\documentclass[preprints,article,accept,pdftex,moreauthors]{Definitions/mdpi} 

\firstpage{1} 
\pubvolume{1}
\issuenum{1}
\articlenumber{0}
\pubyear{2023}
\copyrightyear{2023}
\datereceived{ } 
\daterevised{ } % Comment out if no revised date
\dateaccepted{ } 
\datepublished{ } 
\hreflink{https://doi.org/} % If needed use \linebreak
\Title{The Infrared Extinction Law in the Ophiuchus Molecular Cloud based on UKIDSS and Spitzer}

\TitleCitation{The IR Extinction in the Ophiuchus Cloud}

\Author{Jun Li $^{1,*}$\orcidA{} and Xi Chen $^{1,*}$\orcidB{}}  %and Firstname Lastname $^{2,}$*}

\AuthorNames{Firstname Lastname, Firstname Lastname and Firstname Lastname}

\AuthorCitation{Li, J.; Chen, X.}%; Lastname, F.; Lastname, F.}
\address{%
$^{1}$ \quad Center for Astrophysics, Guangzhou University, Guangzhou 510006, China}

\corres{Correspondence: lijun@gzhu.edu.cn (J. L.); chenxi@gzhu.edu.cn (X. C.) }

\abstract{Investigating the extinction properties in dense molecular clouds is of significant importance for understanding the behavior of interstellar dust and its impact on observations. In this study, we comprehensively examined the extinction law in the Ophiuchus cloud across a wavelength range from 0.8\,$\mu$m to 8\,$\mu$m. To achieve this, we analyzed NIR and MIR data obtained from the UKIDSS GCS and the Spitzer c2d survey, respectively. By fitting a series of color-color diagrams, we determined color-excess ratios $E_{J-\lambda}/E_{J-K}$ for seven passbands. These ratios were then directly converted to derive the relative extinction law $A_\lambda/A_K$. Our findings demonstrate that the Ophiuchus cloud exhibits a characteristic of flat MIR extinction, consistent with previous studies. Additionally, our results reveal variations in the extinction law with extinction depth, indicating a flatter trend from the NIR to MIR bands as extinction increases. Notably, our analysis reveals no significant difference in the MIR extinction law among the four dark clouds: L1712, L1689, L1709, and L1688. However, distinct variations were observed in the extinction law for regions outside the dark clouds, specifically L1688N and L1688W. These regions displayed lower color-excess ratios $E_{J-\lambda}/E_{J-K}$ in the Spitzer/IRAC bands. This observation lends support to the dust growth occurring in the dense regions of the Ophiuchus cloud.}


\begin{document}

%%%%%%%%%%%%%%%%%%%%%%%%%%%%%%%%%%%%%%%%%%
%\setcounter{section}{-1} %% Remove this when starting to work on the template.
%\section{How to Use this Template}

%The template details the sections that can be used in a manuscript. Note that the order and names of article sections may differ from the requirements of the journal (e.g., the positioning of the Materials and Methods section). Please check the instructions on the authors' page of the journal to verify the correct order and names. For any questions, please contact the editorial office of the journal or support@mdpi.com. For LaTeX-related questions please contact latex@mdpi.com.%\endnote{This is an endnote.} % To use endnotes, please un-comment \printendnotes below (before References). Only journal Laws uses \footnote.

% The order of the section titles is different for some journals. Please refer to the "Instructions for Authors” on the journal homepage.

\section{Introduction}\label{sec:intro}

Interstellar dust is a crucial constituent of the interstellar medium (ISM). The physical characteristics of interstellar dust, including its composition, size distribution, and shape, can exhibit significant variations across different interstellar environments\cite{draine2003}. Numerous detailed dust models have been developed to accurately replicate the properties of dust in the diffuse medium\cite{Zubko2004,Compiegne2011,Hensley2021}. However, with an increase in environmental density, dust grains undergo two pivotal evolutionary processes that modify their physical properties. These processes include coagulation, where dust particles collide and merge to form larger grains, and condensation, where gaseous molecules adhere to the surface of dust grains, leading to the formation of ice mantles\cite{Whittet1998,ossenkopf1994,ormel2011}. However, the nature of dust evolution in dense environments remains poorly understood. Gaining a comprehensive understanding of the properties of interstellar dust in dense environments is essential for unraveling the formation and evolution of stars and their planetary systems\cite{Lada1994,Lombardi2001,Alves2001}. 

The observational characteristic that characterizes dust properties is the extinction law, which depicts the wavelength-dependent variation of extinction. The extinction law is commonly employed to infer dust properties, including size distribution and composition\cite{Fitzpatrick1999}. Over the past few decades, the extinction law in the optical and ultraviolet bands has proven valuable in investigating the diffuse interstellar medium, allowing for its characterization through a single parameter (i.e., $R_V=A_V/E(B-V)=A_V/(A_B-A_V)$)\cite{Fitzpatrick1999,cardelli1989}. In the near-infrared (NIR) band, spanning 1-3 $\mu\rm m$, the extinction law has been found to exhibit an approximately universal power-law form $A_\lambda \propto\lambda^{-\alpha}$\cite{wang2014,Schultheis2015}. The power-law index $\alpha$ may vary from a sightline to another in the range of 1.6-2.4 (see the review by Matsunaga et al.\cite{matsunaga2018}).

However, the extinction law in the mid-infrared (MIR) band, spanning 3-8 $\mu$m, lacks a comprehensive description or explanation. The measurement of the extinction law is generally based on the estimation of the entire molecular cloud, but the density of the environment within the same cloud is not uniform. Dust grains in denser environments are more likely to undergo growth, altering their size and composition, thus introducing variations in the infrared extinction law. Nevertheless, the variability of the infrared extinction law with respect to environmental conditions (e.g., temperature and density) remains inadequately investigated. Early studies suggested that the MIR extinction law did not show significant variability\cite{roman2007ApJ,Gao2009}. Subsequent studies revealed that in dense interstellar environments, the MIR extinction law tends to exhibit a flat behavior, aligning with Weingartner \& Draine\cite{weingartner2001} (hereafter WD01) $R_V=5.5$ model. However, Cambr{\'e}sy et al.\cite{cambresy2011} found a "transition" occurs in the MIR extinction law when $A_V>20$ mag, displaying a flatter behavior compared to WD01 $R_V=5.5$ model. Dust growth is suggested to account for this flattened MIR extinction, requiring the dust size to reach approximately several microns\cite{wang2015a,wang2015b}. Ascenso et al.\cite{ascenso2013} found that the MIR extinction curves of two dark cloud cores in Pipe Nebule exhibit a flatter trend compared to the extinction curves predicted by the WD01 dust model, although there exist some disparities between them. In conclusion, notable uncertainties persist in the MIR extinction law within dense environments, primarily attributable to observational constraints concerning accuracy and depth. The question of its universality or its dependence on extinction depth remains a subject of debate.

In order to investigate variations in the extinction law, we chose the nearby Ophiuchus cloud, which exhibits active star formation. The Ophiuchus cloud is located at a distance of approximately 135 pc\cite{Mamajek2008} and is situated at a high galactic latitude. It shows that no contamination by background dark clouds. Consequently, we can consider all stars within the region as background stars. The Ophiuchus cloud is also well-known for its diverse range of star-forming environments, including dense dark clouds and translucent clouds. This characteristic enables us to study possible variations of the extinction law within a single cloud.

In this work, we combine deep NIR and MIR data to study the extinction law in the Ophiuchus cloud. In section \ref{sec:data}, we present the infrared data we used in this paper. Section \ref{sec:method} describes the methods and analysis for obtaining the IR extinction law. Section \ref{sec:result} presents the results and discussions. We make a summary in Section \ref{sec:sum}.

%%%%%%%%%%%%%%%%%%%%%%%%%%%%%%%%%%%%%%%%%%
\section{Observations}\label{sec:data}

We obtained the NIR and MIR observations of Ophiuchus cloud from two catalogs: the United Kingdom Infrared Deep Sky Survey (UKIDSS) Galactic Cluster Survey (GCS) and the Spitzer Cores to Disks (c2d) Project.

\subsection{The UKIDSS GCS}

The UKIDSS GCS is a comprehensive survey aimed at providing homogenous data on a number of open star clusters, star forming regions, and associations\cite{Casewell2013}. This survey utilized the United Kingdom Infrared Telescope (UKIRT) Wide Field Camera (WFCAM) and observed in the $ZYJHK$ bands\cite{Lawrence2007}. The effective wavelengths and full-widths at half-maximum (FWHM) of the UKIDSS filters are presented in Table \ref{tab:1}. The data from the UKIDSS GCS can be accessed through the WFCAM Science Archive \footnote{http://surveys.roe.ac.uk/wsa}. The JHK filters of UKIDSS have significantly deeper detection limits, approximately 3.5 mag deeper than the JHKs bands of the Two Micron All Sky Survey (2MASS), which enables the penatration into regions with up to $A_V\approx40$ mag. In this study, the observation from UKIDSS GCS covers a area of about 6$\times$4 square degrees towards the Ophiuchus cloud. Details regarding the photometric system and calibration can be found in Hewett et al.\cite{Hewett2006} and Hodgkin et al.\cite{Hodgkin2009}, respectively.

\subsection{Spitzer c2d Project}

The Spitzer c2d project is an extensive survey focused on star-forming regions within the Milky Way\cite{evans2003}. This survey employed the Infrared Array Camera (IRAC) and Multiband Imaging Photometer (MIPS) instruments onboard the Spitzer Space Telescope The c2d survey in the Ophiuchus field encompasses an area of approximately 8 square degrees. The dataset comprises imaging observations in four IRAC bands at [3.6], [4.5], [5.8], and [8.0] $\mu$m, as well as in two MIPS bands at [24] and [70] $\mu$m. We only use the Spitzer/IRAC bands in this work. The data from Spitzer c2d project is available in the archive of the Spitzer Science Center's website\footnote{https://irsa.ipac.caltech.edu/data/SPITZER/C2D/}.

\begin{figure}[H]
\includegraphics[width=14 cm]{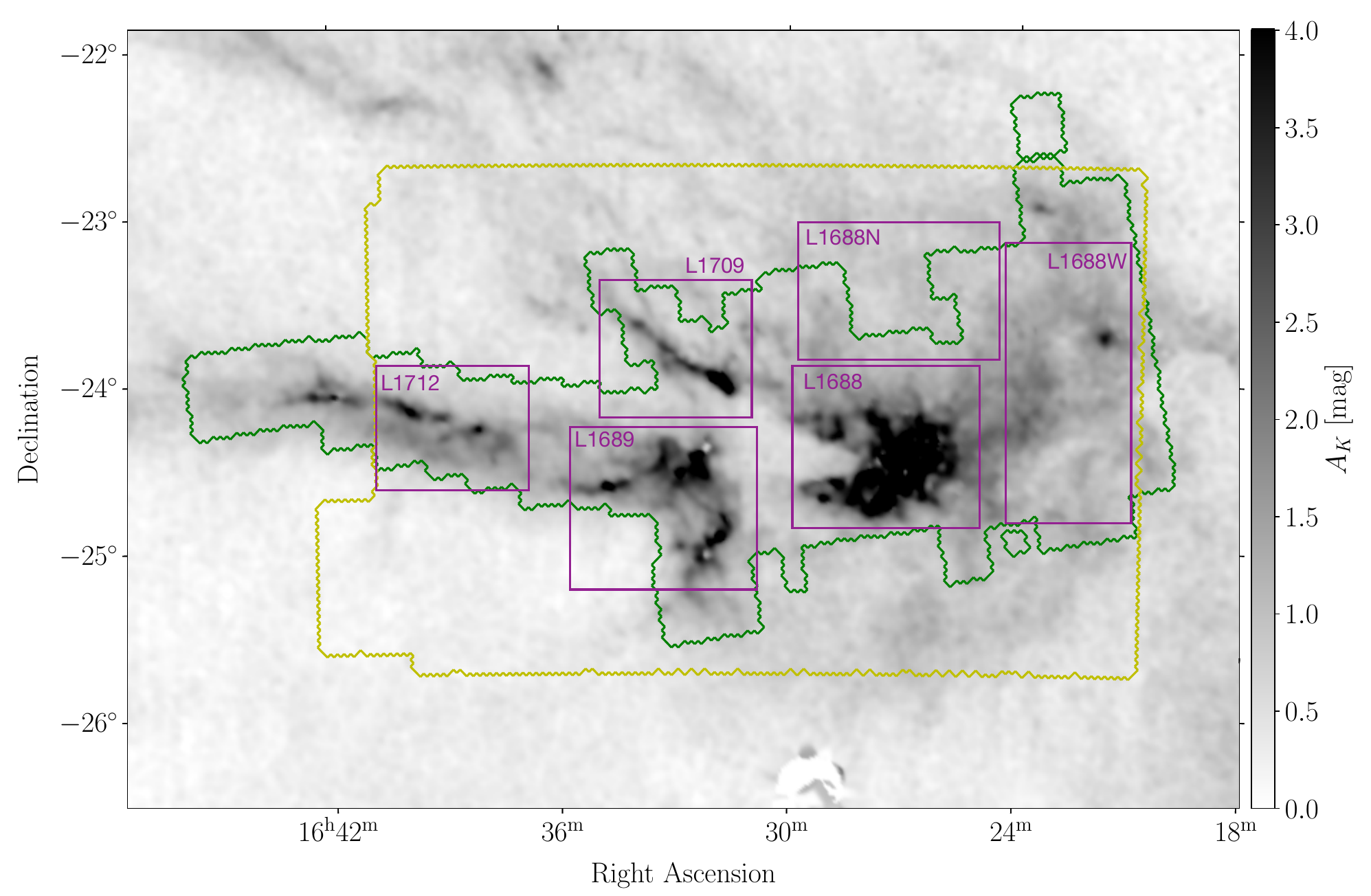}
\caption{The extinction $A_K$ map of Ophiuchus Cloud determined by Juvela \& Montillaud\cite{juvela2016} using the 2MASS survey. The resolution of \textbf{the} $A_K$ map is $\sim$3.0 arcmin. The irregular outlines in green and yellow indicate the spatial coverage of the Spitzer and UKIDSS observations, respectively. The entire region is subdivided into six distinct sub-regions, delineated by solid-line rectangles, namely L1712, L1689, L1709, L1688, L1688N, and L1688W. \label{fig:ak-map}}
\end{figure}

Figure \ref{fig:ak-map} presents the observation coverages of UKIDSS and Spitzer, overlaid on the $A_K$ extinction map calculated by Juvela \& Montillaud\cite{juvela2016} using 2MASS survey. In a related study, Lombardi et al.\cite{Lombardi2008} presented the first $A_K$ extinction map of Ophiuchus based on the 2MASS data by using the Near-Infrared Color Excess Revisited (NICER\cite{Lombardi2001}) method. Additionally, Figure \ref{fig:ak-map} displays the positions of six specific sub-regions, namely L1712, L1689, L1709, L1688, L1688N, and L1688W, which will be further explored and discussed in Section \ref{subsec:varia}. 

%\unskip

\subsection{Control of the Data Quality}

To determine the extinction law of Ophiuchus cloud in both NIR and MIR bands, we utilized the TOPCAT\cite{Taylor2005} software to cross-match the UKIDSS and Spitzer catalogs described above. The matching radius was set to be 1 arcsecond to obtain sources with simultaneous observations. For the UKIDSS data, We first adopt a strict cut on uncertainties for UKIDSS $ZYJHK$ bands by requiring photometric magnitude errors smaller than 0.1 mag. Furthermore,  the values of `pstar' flag were restricted to be greater than 0.99, to eliminate non-stellar sources such as galaxies and extended sources. As a result, the obtained dataset exhibits limiting magnitudes of 19.4, 17.8, and 16.9 mag in $JHK$ bands, respectively.

For the Spitzer data, we applied quality criteria to select sources classified as `star' according to the `OType' flag, and we retained only those sources with a signal-to-noise ratio (SNR) greater than 5. The resulting limiting magnitudes for the four IRAC bands are 16.6, 15.9, 15.3, and 14.8 mag. Moreover, we excluded YSOs within the cloud as identified by Grasser et al.\cite{grasser2021}. Number counts of all bands under data quality control are tabulated in Table \ref{tab:1}.

\begin{table}[H] 
\caption{Basic properties of all photometric bands used in this work, as well as the detection limits and the number of sources in our merged catalogue for the Ophiuchus cloud.\label{tab:1}}
%\newcolumntype{c}{>{\centering\arraybackslash}X}
\begin{tabularx}{\textwidth}{CCCCC}
\toprule 
\textbf{Bands}	& $\lambda_{\rm eff}^0$ ($\mu \rm m$) & FWHM ($\mu \rm m$) & Limits (mag) & Number \\
\midrule
$Z$     & 0.882  & 0.092  & 20.1  & 15743 \\
$Y$     & 1.031  & 0.103  & 19.5  & 16714 \\
$J$     & 1.248  & 0.158  & 19.4  & 18078 \\
$H$     & 1.627  & 0.295  & 17.8  & 17494 \\
$K$     & 2.188  & 0.351  & 16.9  & 18456 \\
$[3.6]$ & 3.508  & 0.743  & 16.6  & 18601 \\
$[4.5]$ & 4.414  & 1.010  & 15.9  & 18610 \\
$[5.8]$ & 5.648  & 1.391  & 15.3  & 12104 \\
$[8.0]$ & 7.592  & 2.831  & 14.8  & 6751  \\
\bottomrule
\end{tabularx}
%\textsuperscript{1}
%\noindent{\footnotesize{\textsuperscript{1} Tables may have a footer.}}
\end{table}

%%%%%%%%%%%%%%%%%%%%%%%%%%%%%%%%%%%%%%%%%%
\section{Methods and Analysis}\label{sec:method}

\subsection{Determination of the Extinction Law}\label{subsec:met}

In this study, we investigate the wavelength-dependent extinction law using a technique similar to that employed by Indebetouw et al.\cite{indebetouw2005}. However, we deviate from their approach by measuring the color excess ratios $\beta_\lambda\equiv E_{J-\lambda}/E_{J-K}$ instead of $E_{\lambda-K}/E_{J-K}$. Initially, we determine the color excess ratios $E_{J-\lambda}/E_{J-K}$ by fitting the loci of the stellar population on the color-color diagram (i.e. $J-
\lambda$ vs. $J-K$). Subsequently, we determine the relative extinction law from $\beta_\lambda$ and $A_J/A_K$:
\begin{equation}\label{equ:1}
    A_\lambda/A_K = (1-A_J/A_K)\beta_\lambda+A_J/A_K
\end{equation}
The value of $A_J/A_K$ is often adopted to be $2.5\pm 0.15$, as measured by Indebetouw et al. \cite{indebetouw2005} using the 2MASS survey. However, it is crucial to question this assumption. To address this, we recalculate $A_J/A_K$ using the color excess ratio $E_{J-H}/E_{J-K}$ obtained from the $J-H$ vs. $J-K$ diagram. Although the UKIDSS photometry is calibrated relative to 2MASS, the systematic biases were found in the photometry between UKIDSS and 2MASS\cite{cambresy2011}. This recalculation is necessary due to the considerable differences in filters between UKIDSS and 2MASS. Furthermore, it is important to note that the NIR extinction law may exhibit variations depending on the interstellar environment. We assume a power-law form for the extinction law in the NIR $JHK$ bands, given by $A_\lambda\propto\lambda^{-\alpha}$. Consequently, the color excess ratio can be used to derive the value of the power-law index $\alpha$ using the following relationship:
\begin{equation}\label{equ:2}
    \frac{E_{J-H}}{E_{J-K}}=\frac{(\frac{\lambda_K}{\lambda_J})^\alpha - (\frac{\lambda_K}{\lambda_H})^\alpha}{(\frac{\lambda_K}{\lambda_J})^\alpha-1}
\end{equation}
where $\lambda_J$, $\lambda_H$, and $\lambda_K$ are the effective wavelengths of the UKIDSS $JHK$ bands, respectively. Hence, our initial step involves deriving $\alpha$ and the relative extinction in the NIR, specifically $A_J/A_K$ and $A_H/A_K$, for each sample using the UKIDSS data.

The estimation of color excess ratios $E_{J-\lambda}/E_{J-K}$ is performed by constructing a color-color diagram, specifically $J-\lambda$ vs. $J-K$. The positioning of star populations on the color-color diagram is influenced by multiple factors, including their intrinsic colors, extinction, and photometric errors. In this investigation, the Ophiuchus cloud exhibits significant extinction in the $K$ band, with $A_K$ exceeding 4 mag (see Figure \ref{fig:ak-map}). Moreover, the observed $J-K$ spans a wide range, with $\Delta(J-K)$ exceeding 7 mag. The intrinsic color indices typically display a dispersion of approximately 0.2 mag. Additionally, we have exclusively selected sources with photometric errors below 0.1 mag for our analysis. Consequently, the dominant factor shifting the stars along the reddening vector in the color-color diagram is the effect of extinction.

When conducting a linear fit to the $J-\lambda$ vs. $J-K$ color-color diagram, we apply a $3\sigma$ removal technique to identify and exclude outliers in a single iteration. Additionally, we exclude sources with $J-K<0.65$ mag from the analysis.

\subsection{Non-Linearity of Extinction Coefficients}

\begin{figure}[H]
\centering
\includegraphics[width=13.5 cm]{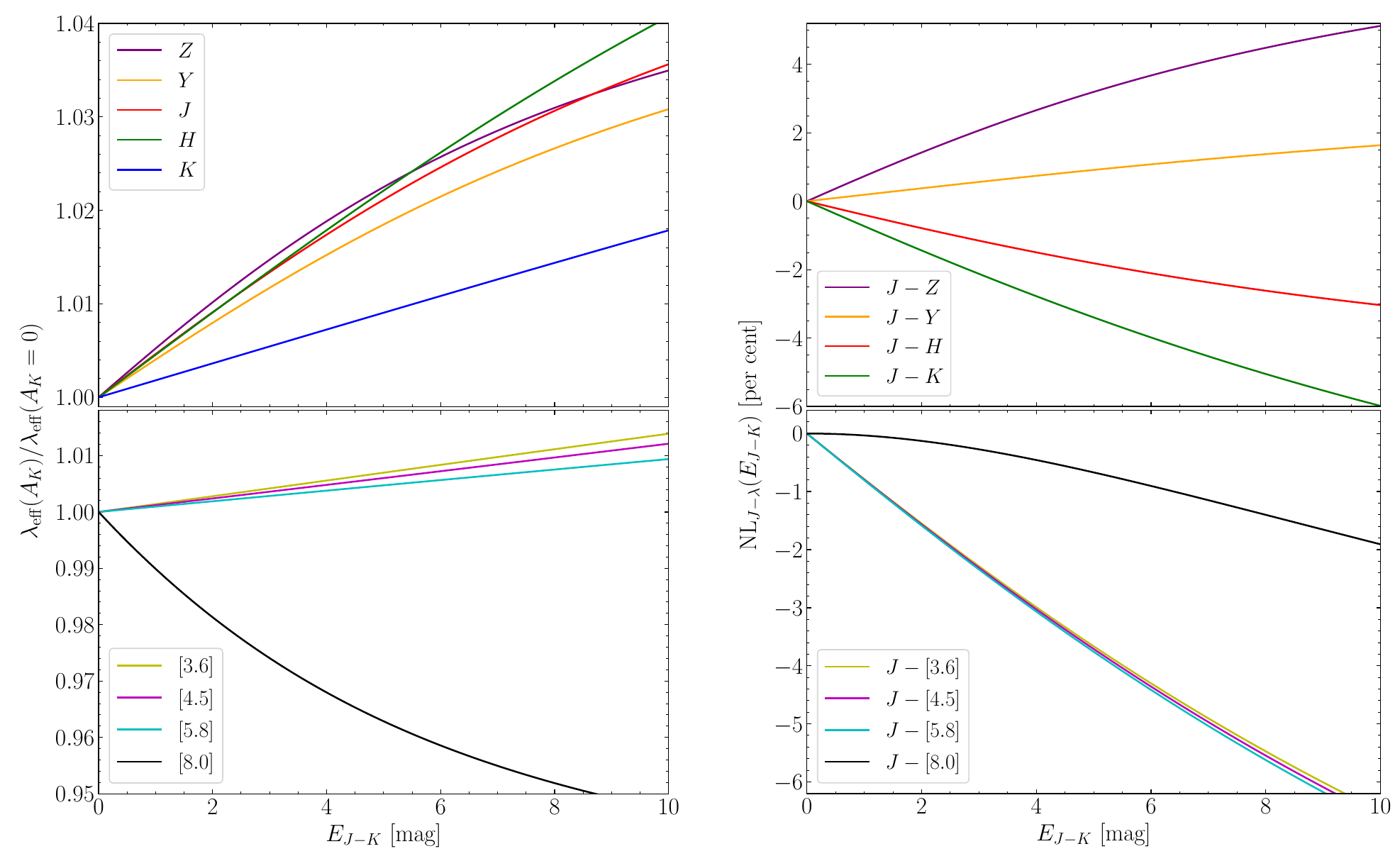}
\caption{\emph{Left panels:} The ratio of effective wavelength at $A_K$ to static wavelength at $A_K=0$ ($\lambda_{\rm eff}(A_K)/\lambda_{\rm eff}(A_K=0)$) as a function of $E_{J-K}$. \emph{Right panels:} The non-linear correction factor (${\rm NL}_{J-\lambda}(E_{J-K})$) as a function of $E_{J-K}$. The lines in different colors represent the filters of UKIDSS and Spitzer. We adopted a stellar spectrum of a typical K0III-type star and extinction curve of WD01 $R_V$=5.5 model. \label{fig:NLc}} 
\end{figure}

It is widely acknowledged in the literature that the effective wavelengths of filters undergo a gradual shift towards longer wavelengths\cite{stead2009,wang2019,meingast2018}. Consequently, the non-linearity in the extinction calculation (i.e., Equation \ref{equ:1} and \ref{equ:2}) becomes significant. In previous studies, this non-linear effect has been commonly regarded as negligible. However, with the increasing depth of extinction, this effect becomes more prominent and cannot be disregarded. The effective wavelength of a filter is determined by convolving the stellar spectra with the filter transmission curve:
\begin{equation}
    \lambda_{\rm eff}=\frac{\int \lambda F(\lambda)T(\lambda)e^{-A(\lambda)/1.086}{\rm d}\lambda}{\int F(\lambda)T(\lambda)e^{-A(\lambda)/1.086}{\rm d}\lambda}
\end{equation}
where $F_\lambda$ represents the intrinsic flux of the stellar spectra, and $T(\lambda)$ denotes the relative response function of the filter. The effective wavelength when the extinction $A_V=0$ is defined as the static effective wavelength $\lambda_{\rm eff}^0$, which is listed in Table \ref{tab:1}. However, the effective wavelength is subject to shifting relative to static wavelength due to extinction $A(\lambda)$. To analyze the "shifting effect," we utilize a K0III-type stellar spectrum from Castelli \& Kurucz\cite{castelli2003} with an effective temperature of 4750\,K, $logg$=2.5, and solar metallicity. We assume the extinction law of the WD01 $R_V=5.5$ model. The left panels in Figure \ref{fig:NLc} depict the ratio of the effective wavelength $\lambda_{\rm eff}$ at $A_K$ to the static wavelength $\lambda_{\rm eff}(A_K=0)$ as a function of the reddening $E_{J-K}$. With the exception of the effective wavelength of the IRAC [8.0] band, which shifts towards shorter wavelengths, the effective wavelengths of all other bands shift towards longer wavelengths. This difference can be attributed to the 9.8 $\mu$m extinction feature caused by silicate dust. The effective wavelengths of most bands have relative changes of a few per cent over the range of considered reddening $E_{J-K}<10$ mag.

The extinction $A_{\lambda_x}$ in the $\lambda_x$ band is calculated using the following formula:
\begin{equation}
    A_{\lambda_x}=-2.5{\rm log_{10}} \frac{\int \lambda T_x(\lambda) F(\lambda) 10^{-0.4A(\lambda)}{\rm d}\lambda}{\int\lambda T_x(\lambda) F(\lambda) {\rm d}\lambda}
\end{equation}
The color excess $E_{\lambda_x-\lambda_y}$ between two photometric systems is equal to $A_{\lambda_x} - A_{\lambda_y}$. The shift effect of the effective wavelength can be corrected by applying an extinction correction to the observed data, which depends on the extinction law and the amount of reddening along the line of sight. According to Sanders et al.\cite{sanders2022}, the non-linear correction factor for the color $J-\lambda$ (denoted as ${\rm NL}_{J-\lambda}(E_{J-K})$) as a function of reddening $E_{J-K}$ is defined as follows:
\begin{equation}
    {\rm NL}_{J-\lambda}(E_{J-K})=E_{J-\lambda}/E_{J-K}-E_{J-\lambda}/E_{J-K}|_{A_\lambda\rightarrow 0}
\end{equation}

We apply a correction for the typical non-linear effects by considering a K0III-type stellar model described above, combined with the WD01 $R_V$=5.5 extinction law. The observed colors of stars are corrected using the formula $(J-\lambda)\leftarrow(J-\lambda)-E_{J-K}\times{\rm NL}_{J-\lambda}(E_{J-K})$, where we approximate $E_{J-K}$ as $(J-K)-0.65$. The right panels of Figure \ref{fig:NLc} illustrate the variation of the non-linear correction factors ${\rm NL}_{J-\lambda}(E_{J-K})$ for UKIDSS and Spitzer/IRAC bands as a function of $E_{J-K}$.

%\unskip

\section{Results and Discussions}\label{sec:result}

\subsection{Determination of $A_J/A_K$ }\label{subsec:a-j}

\begin{figure}[H]
\centering
\includegraphics[width=13 cm]
{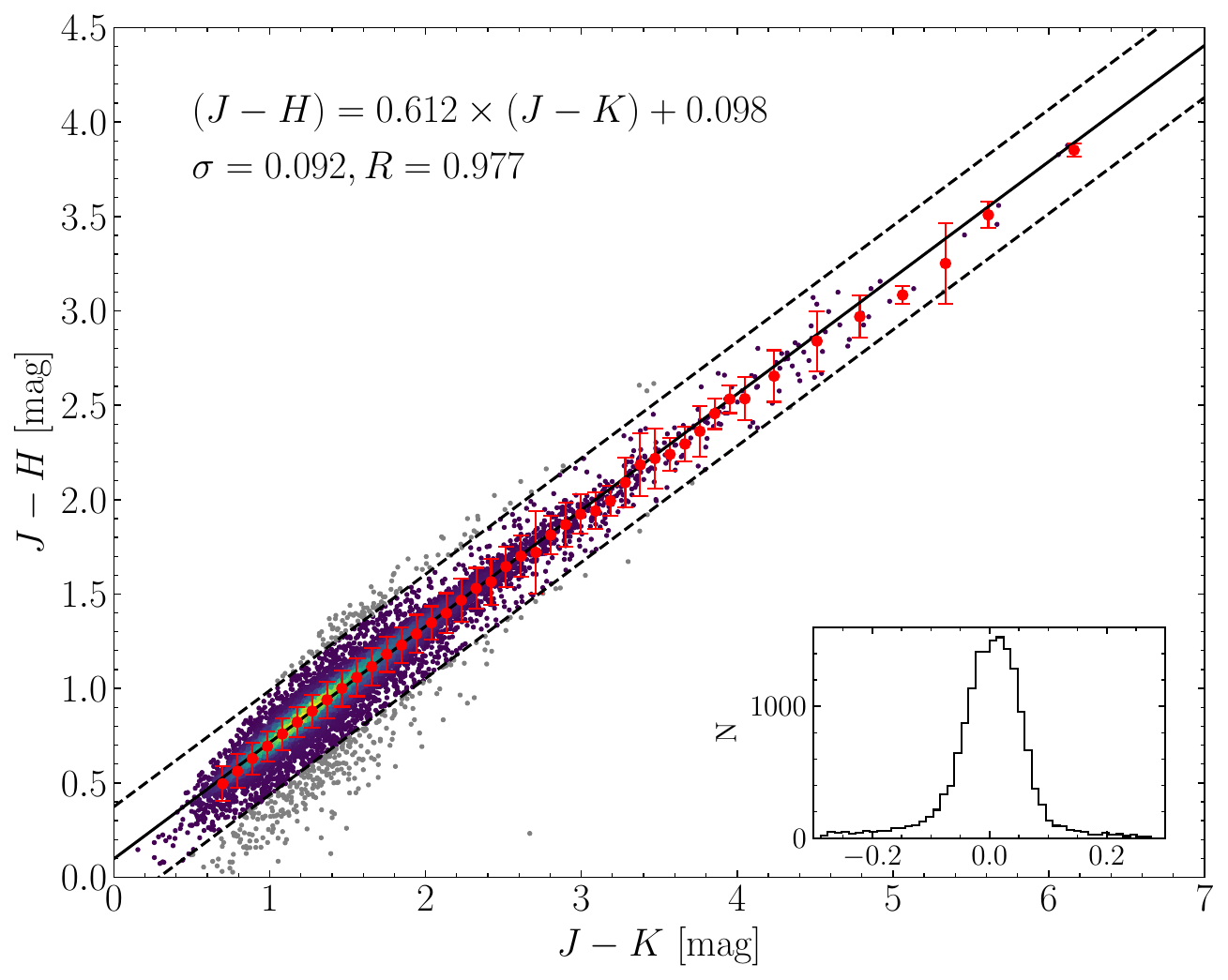}
\caption{The $J-H$ vs. $J-K$ diagram for sources observed by both Spitzer and UKIDSS in the Ophiuchus cloud, where the color-scale represent the number density. The black solid line denotes the best linear fit to the points after a 3$\sigma$ removal, whose parameters are also shown in the figure.
The $3\sigma$ range is shown in dashed lines, and the outliers are shown in grey. The inset is the histogram of the residuals between best-fit line and observed points. \label{fig:jh-jk}}
\end{figure}

As mentioned in Section \ref{subsec:met}, in order to convert the color excess ratio $E_{J-\lambda}/E_{J-K}$ into the relative extinction $A_\lambda/A_K$, it is necessary to measure the NIR extinction law $A_J/A_K$. Utilizing the $JHK$ photometric data from UKIDSS, we initially performed a linear fit on the $J-H$ vs. $J-K$ color-color diagram, yielding $(J-H)\approx 0.612\times (J-K)+0.098$. As depicted in Figure \ref{fig:jh-jk}, it is evident that the best-fitting line exhibits an excellent linear relationship with the observations, accompanied by minimal error. For comparison, we divided the $J-K$ data set into bins with variable widths. We then calculated the mean and standard deviation of the corresponding $J-H$ values within each bin. It is clearly seen that these mean points exhibit a remarkable alignment with the best-fit line by fitting the observed data directly. The distribution of residuals in the histogram further confirms that the fitting results do not exhibit significant systematic biases in relation to the data (see the inset in Figure \ref{fig:jh-jk}).

Previous studies have investigated the NIR extinction law in the Ophiuchus cloud by employing different methodologies and photometric systems\cite{Harris1978,kenyon1998,naoi2006}. Kenyon et al.\cite{kenyon1998} obtained a value of $E_{J-H}/E_{H-K}=1.57$, while Naoi et al.\cite{naoi2006} reported a range of $E_{J-H}/E_{H-K}=1.60-1.69$, which exhibited slight variations depending on the location within the cloud. In our analysis using UKIDSS data in the Ophiuchus cloud, we derived a value of approximately $E_{J-H}/E_{H-K}\approx1.58$, demonstrating excellent agreement with their findings.

According to the conversion in Equation \ref{equ:2}, using the derived $E_{J-H}/E_{J-K}=0.612\pm0.001$ and the static effective wavelengths for the UKIDSS $JHK$ bands ($\lambda_J=$1.248 $\mu$m, $\lambda_H=$1.627 $\mu$m, and $\lambda_K=$2.188 $\mu$m, as shown in Table \ref{tab:1}), we yield a power-law index of $\alpha=2.028\pm0.015$. Consequently, the extinctions for the $J$ and $H$ bands are determined to be $A_J/A_K=3.123\pm0.027$ and $A_H/A_K=1.824\pm0.008$, respectively. Previous studies have reported varying values of $A_J/A_K$, which depend on the interstellar environment and the specific photometric system used in the telescopes. These studies include Rieke \& Lebofsky\cite{rieke1985} who obtained $A_J/A_K$ = 2.52 towards the Galactic Center using the IRTF, Indebetouw et al.\cite{indebetouw2005} who reported $A_J/A_K$ = 2.50 by averaging two different sight-lines using the 2MASS survey, Nishiyama et al.\cite{nishiyama2009} who found $A_J/A_K$ = 3.02 towards the Galactic Center using the 2MASS survey, Fritz et al.\cite{fritz2011} who derived $A_J/A_K$ = 3.07 towards the Galactic Center using hydrogen lines, Gordon et al.\cite{Gordon2021} who derived the $A_J/A_K$ = 2.53 for the average diffuse extinction, and Sanders et al.\cite{sanders2022} who obtained $A_J/A_K$ = 3.23 towards the inner region of the Milky Way using the VVV survey. In our study, we employ the UKIRT photometric system, and the derived value of $A_J/A_K=3.12\pm0.03$ will be used to calculate the relative extinction $A_\lambda/A_K$ in other bands.

%\unskip

\subsection{Determination of $A_\lambda/A_K$}\label{subsec:a-lambda}

\begin{figure}[H]
\centering
\includegraphics[width=13.6 cm]{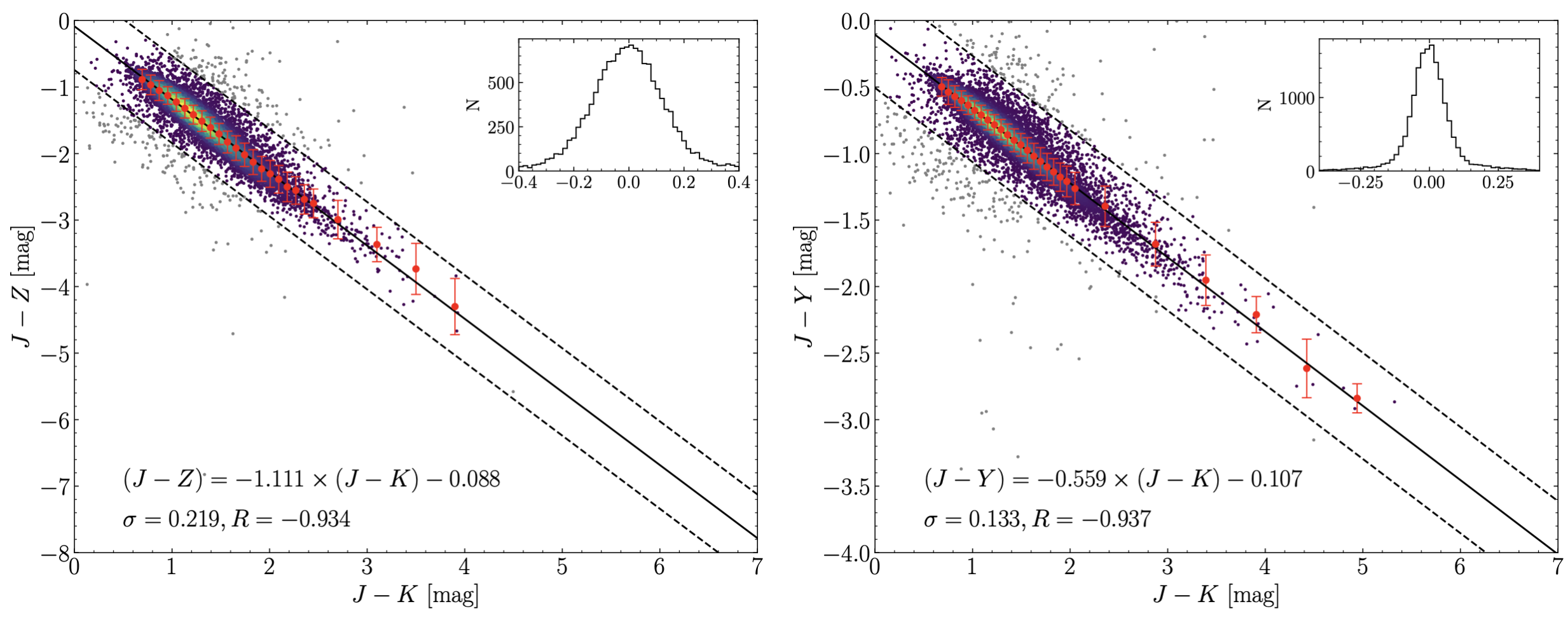}
\caption{Color vs. color diagrams $J-\lambda$ vs. $J-K$ for UKIDSS $Z$ and $Y$ bands. The conventions of symbols and lines are the same as in Figure \ref{fig:jh-jk}. \label{fig:jzy-jk}}
\end{figure}

\begin{figure}[H]
\centering
\includegraphics[width=13.8 cm]{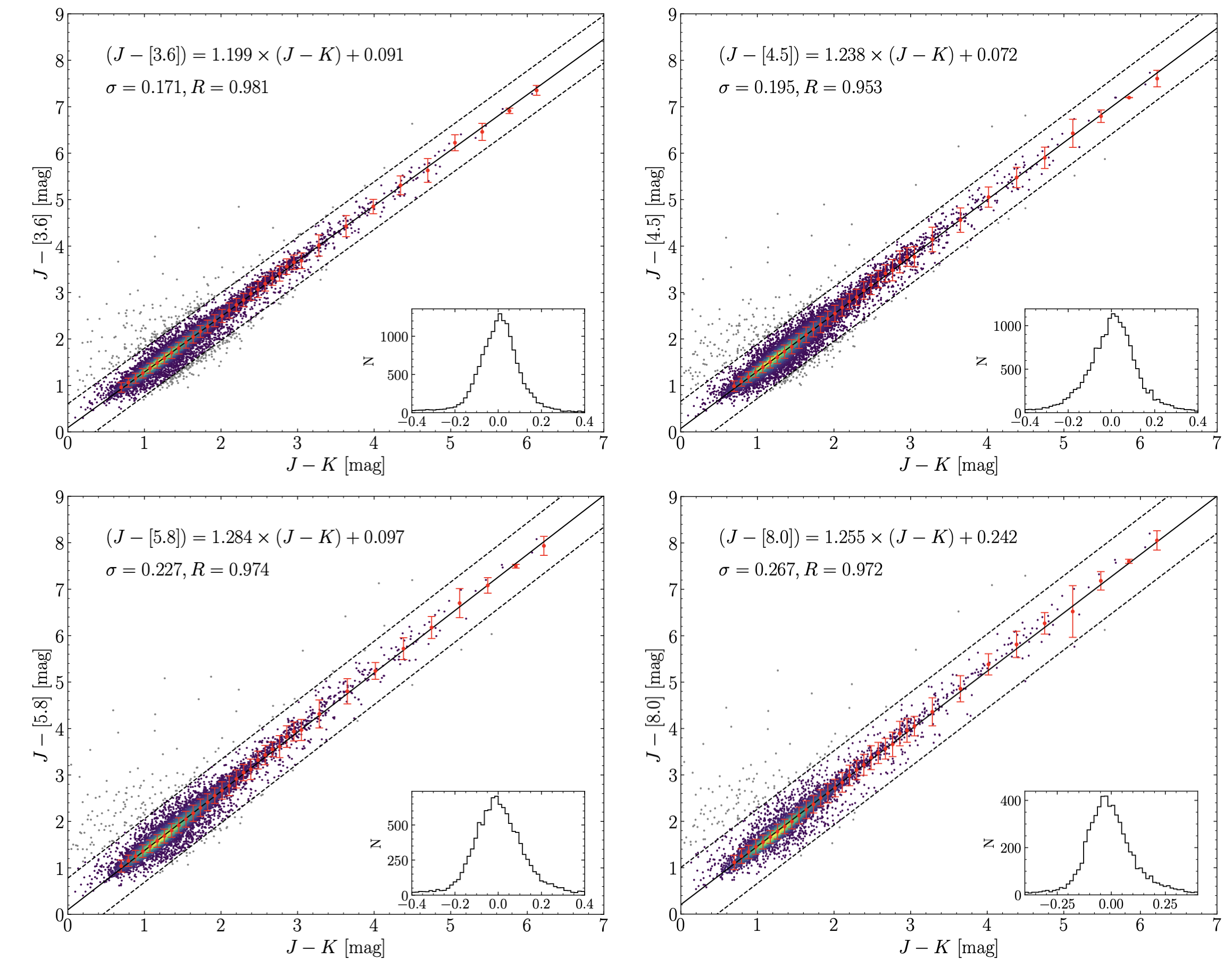}
\caption{Color vs. color diagrams $J-\lambda$ vs. $J-K$ for Spitzer/IRAC [3.6], [4.5], [5.8] and [8.0] bands. The conventions of symbols and lines are the same as in Figure \ref{fig:jh-jk}.\label{fig:jirac-jk}}
\end{figure}

To investigate the overall extinction law of the Ophiuchus cloud, we conducted an analysis of the color-color diagrams depicting the relationship between $J-\lambda$ and $J-K$ across various bands, as illustrated in Figures \ref{fig:jzy-jk} and \ref{fig:jirac-jk}. The objective was to determine the slope of the color-color relations through linear fitting, which represents the color excess ratio denoted as $\beta_\lambda = E_{J-\lambda}/E_{J-K}$. In Figure \ref{fig:jirac-jk}, noticeable photometric outliers were observed, characterized by unconventional colors, such as stars exhibiting $(J-K)$ values ranging from 0 to 2 mag but displaying significantly large $(J-\lambda)$ values. These outliers likely correspond to sources with infrared excess resulting from circumstellar dust. To address this issue, a 3$\sigma$ removal procedure was applied in a single iteration. Initially, all data points were fitted, and the residuals between the points and the best-fit line were calculated. Subsequently, data points beyond 3$\sigma$ of the standard deviation of residuals were excluded. As depicted in Figures \ref{fig:jzy-jk} and \ref{fig:jirac-jk}, the gray points represent the outliers that were not considered in the final fitting process. Following this removal step, the remaining sources were refitted to obtain the definitive fitting results. The resulting color excess ratios $\beta_\lambda=E_{J-\lambda}/E_{J-K}$ were tabulated in Table \ref{tab:2}.

By utilizing the derived $E_{J-\lambda}/E_{J-K}$ and $A_J/A_K=3.12\pm0.03$, we employed Equation \ref{equ:1} to calculate the extinction relative to the $K$ band $A_\lambda/A_K$. The calculated values are presented in Table \ref{tab:2}. The uncertainties associated with $A_\lambda/A_K$ were determined by propagating the uncertainties from both $E_{J-\lambda}/E_{J-K}$ and $A_J/A_K$. Additionally, for comparison purposes, we computed the extinction $A_\lambda/A_K$ using $A_J/A_K=2.50\pm0.15$ obtained from Indebetouw et al.\cite{indebetouw2005}. It is important to note that a smaller value of $A_J/A_K$ leads to a flatter $A_\lambda/A_K$ relationship, resulting in larger values of $A_\lambda/A_K$ in the IRAC bands and smaller values of $A_\lambda/A_K$ in the UKIDSS bands.

\unskip

%\unskip

\begin{figure}[H]
\centering
\includegraphics[width=13 cm]{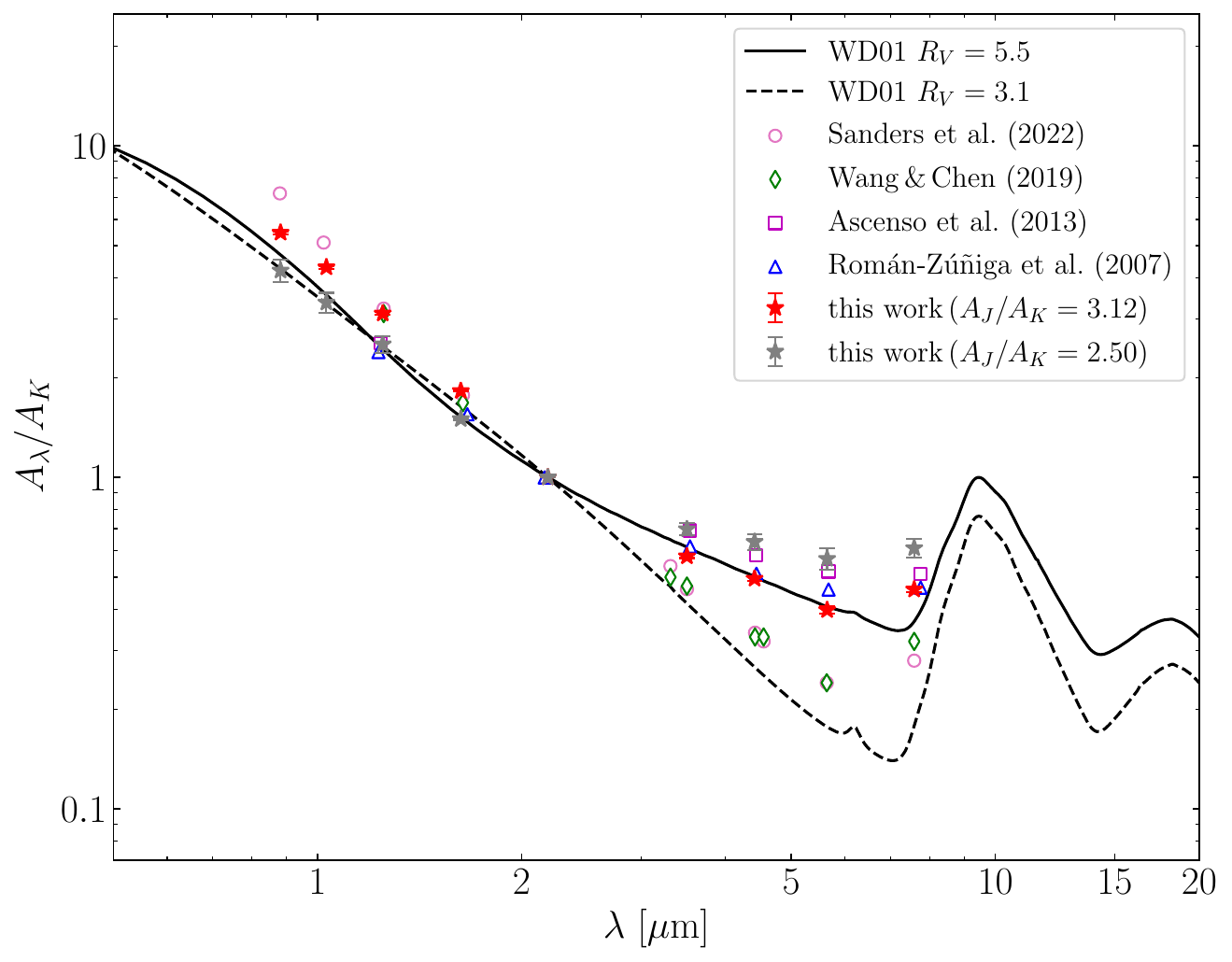}
\caption{Comparison of the extinction laws derived in this work with previous determinations and model predictions. The extinction $A_\lambda/A_K$ determined by adopting $A_J/A_K=3.12$ and $A_J/A_K=2.55$ are shown with red and gray stars, respectively. The extinction laws as predicted from models by WD01 $R_V=3.1$ and $R_V=5.5$ are displayed black solid and dashed lines, respectively. \label{fig:ext-all}}
\end{figure}

Figure \ref{fig:ext-all} presented the extinction law derived for the Ophiuchus cloud in the wavelength range of 0.8-8 $\mu$m. The extinction curves predicted by WD01 models with $R_V=3.1$ and $R_V=5.5$\cite{weingartner2001} and previous measurements\cite{ascenso2013,wang2019,sanders2022} of the infrared extinction law are also shown for comparison. The WD01 models with $R_V$=3.1 and $R_V$=5.5 are generally used to represent the extinction laws in the diffuse and dense ISM, respectively. In the NIR ($0.8-3\,\mu\rm m$), the predicted extinction laws from the WD01 models with $R_V=3.1$ and $R_V=5.5$ exhibit minimal differences. The derived $A_\lambda/A_K$ in the NIR also follow a power-law form and are slightly steeper than the WD01 models. Our results lie between the WD01 models and those of Sanders et al.\cite{sanders2022}. These findings confirm that the NIR extinction law may generally follow a universal power-law form, although the power index $\alpha$ may vary due to differences in the interstellar environment or photometric systems.

\begin{table}[H] 
\centering
\caption{Results for the color-excess ratio $\beta_\lambda=E_{J-\lambda}/E_{J-K}$, as well as relative extinction $A_\lambda/A_K$ from $A_J/A_K=3.12$ of our adoption and $A_J/A_K=2.50$ of Indebetouw et al.\cite{indebetouw2005}.\label{tab:2}}
%\newcolumntype{C}{>{\centering\arraybackslash}X}
%\begin{tabularx}{\textwidth}{CCCC}
\begin{tabularx}{12cm}{cCCC}
\toprule  %$\sigma_{\textbf{\emph{JHK}}}$
\textbf{Band} & $\beta_\lambda$ & $A_\lambda/A_K$&$A_\lambda/A_K$ \\
\cmidrule(r){3-4}
 &   & ($A_J/A_K=3.12$)  & ($A_J/A_K=2.50$) \\
\midrule
$Z$ & -1.111$\pm$0.003 & 5.475$\pm$0.057 & 4.209$\pm$0.317\\
$Y$ & -0.559$\pm$0.002 & 4.305$\pm$0.042 & 3.370$\pm$0.234\\
$H$ & 0.612$\pm$0.001 & 1.824$\pm$0.008 & 1.550$\pm$0.100 \\
$[3.6]$ &1.199$\pm$0.002 & 0.578$\pm$0.007 & 0.698$\pm$0.030\\
$[4.5]$ & 1.238$\pm$0.003 & 0.495$\pm$0.008 & 0.639$\pm$0.036\\
$[5.8]$ & 1.284$\pm$0.003 & 0.398$\pm$0.010 & 0.568$\pm$0.043\\
$[8.0]$ & 1.255$\pm$0.003 & 0.459$\pm$0.009 & 0.612$\pm$0.039\\
%\cmidrule{2-6}
\bottomrule
\end{tabularx}
%\textsuperscript{1}
%\noindent{\footnotesize{\textsuperscript{1} Tables may have a footer.}}
\end{table}

%\unskip

In the MIR range of 3-8 $\mu\rm m$, the extinction predicted by the WD01 $R_V=5.5$ model is noticeably higher than that of the WD01 $R_V=3.1$ model. By using $A_J/A_K=2.50$ instead of $A_J/A_K=3.12$, the derived values of $A_\lambda/A_K$ show an increase of approximately 33\%, 43\%, 29\%, and 20\% for the [3.6], [4.5], [5.8], and [8.0] bands, respectively. Consequently, the extinction law in the MIR appears to be significantly flatter than the prediction of the WD01 $R_V=5.5$ model, as illustrated in Figure \ref{fig:ext-all}.

Previous studies by Ascenso et al.\cite{ascenso2013} and Rom{\'a}n-Z{\'u}{\~n}iga et al.\cite{roman2007ApJ} have investigated the extinction law in dark cloud cores using the color-excess ratio method. They employed the value $A_H/A_K=1.55$ from Indebetouw et al.\cite{indebetouw2005} and derived a MIR extinction law that was significantly flatter than the WD01 $R_V=5.5$ model. In contrast, Figure \ref{fig:ext-all} in our study demonstrates that our results, based on the adoption of $A_J/A_K=2.50$ from Indebetouw et al.\cite{indebetouw2005}, closely resemble the extinction law obtained by Ascenso et al.\cite{ascenso2013}. It can be observed that when measuring the extinction law using color-excess ratio method, employing different $A_J/A_K$ or $A_H/A_K$ can lead to significantly different relative extinction laws. Overall, by utilizing our derived value of $A_J/A_K$ for Ophiuchus cloud, our findings reveal a notable characteristic of a flat MIR extinction. This measurement closely aligns with the extinction curve predicted by the WD01 $R_V$=5.5 model\cite{weingartner2001} and is in agreement with previous measurements\cite{chapman2009,flaherty}.

\subsection{Dependence on Extinction Depth}\label{subsec:depth}

The observational data utilized in this study cover a wide range of extinction values within the Ophiuchus molecular cloud, spanning approximately $\Delta(J-K)\approx7$ mag. This corresponds to an approximate extinction range of $\Delta A_K\approx4.5$ mag, considering an intrinsic color of $(J-K)_0=0.65$ mag and assuming an extinction law of $A_K/E_{J-K}=0.695$ predicted by the WD01 $R_V=5.5$ model. The extensive coverage enables the investigation of variations in the extinction law within the same cloud as a function of extinction depth. Following the division scheme proposed by Chapman et al.\cite{chapman2009}, the data in this study are categorized into four sub-samples based on extinction: $A_K\leq0.5$, $0<A_K\leq1$, $1<A_K\leq2$, and $A_K>2$. The corresponding observed color ranges are $(J-K)\leq1.37$, $1.37<(J-K)\leq2.09$, $2.09<(J-K)\leq3.53$, and $(J-K)>3.53$. Subsequently, a linear fit is performed for each sub-sample to derive the color excess ratio $E_{J-\lambda}/E_{J-K}$, and the results are presented in Table \ref{tab:3}. It can be observed that, except for the sub-sample with $A_K<0.5$, the variation of $E_{J-H}/E_{J-K}$ with increasing $A_K$ is minimal, remaining approximately around 0.6. However, for the four Spitzer/IRAC bands, there is a noticeable decreasing trend of $E_{J-\lambda}/E_{J-K}$ with increasing $A_K$.

\begin{figure}[H]
\centering
\includegraphics[width=13 cm]{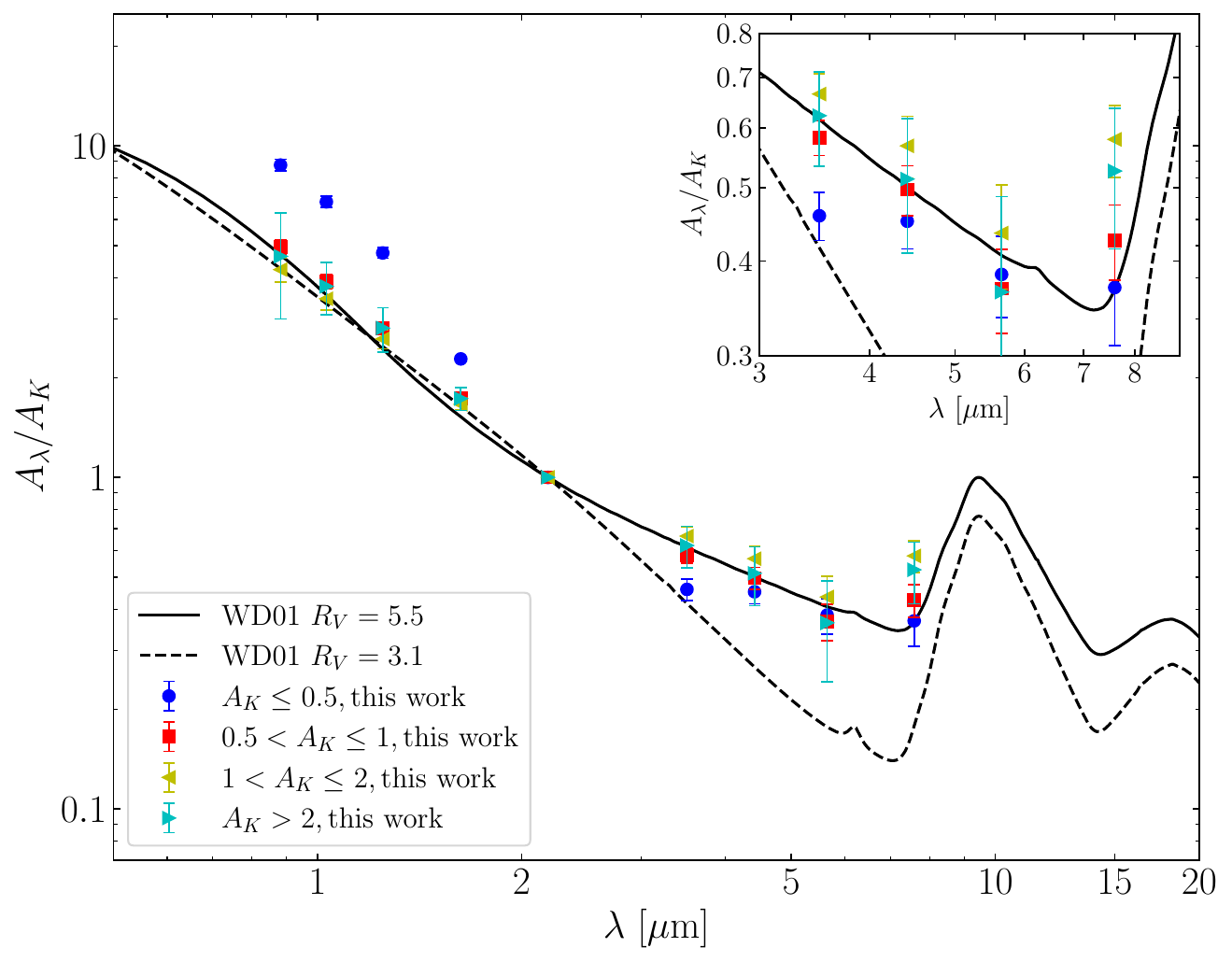}
\caption{Comparison of the extinction $A_\lambda/A_K$ in various $A_K$ ranges. The extinction law as predicted from models by WD01 $R_V=3.1$ and $R_V=5.5$ are displayed black solid and dashed lines, respectively. The inset is the zoom-in view of the MIR extinction. \label{fig:ext-depth} }
\end{figure}

Based on the obtained $E_{J-H}/E_{J-K}$ values for each sub-sample as listed in Table \ref{tab:4}, we can further calculate the power-law index $\alpha$ and $A_J/A_K$ for the NIR extinction law using Equation \ref{equ:2}. By utilizing the derived $A_J/A_K$ values for each sub-sample, we are able to calculate the extinction ratios $A_\lambda/A_K$. These results are presented in Table \ref{tab:4} and visualized in Figure \ref{fig:ext-depth}. It is evident that the extinction law becomes flatter with increasing $A_K$. Specifically, the values of $A_\lambda/A_K$ in the NIR bands decrease, while those in the MIR bands increase with increasing $A_K$. Overall, our findings are consistent with the study conducted by Chapman et al.\cite{chapman2009}, who also observed a similar trend of the extinction law becoming flatter with increasing extinction. However, it should be noted that there are significant uncertainties in the results of the sub-sample with $A_K>2$.

\begin{table}[H] 
\centering
\caption{Results for the color-excess ratio $\beta_\lambda=E_{J-\lambda}/E_{J-K}$ within various ranges of $A_K$. The numbers of sources used for each sub-sample are also listed. \label{tab:3}}
\begin{adjustwidth}{-\extralength}{0cm}
%\newcolumntype{C}{>{\centering\arraybackslash}X}
%\begin{tabularx}{\\fulllength}{CCCC}
\begin{tabularx}{\fulllength}{cCcCcCcCc}
\toprule
Band & \multicolumn{2}{c}{$A_K\leq0.5$}  & \multicolumn{2}{c}{$0.5<A_K\leq1$} & \multicolumn{2}{c}{$1<A_K\leq2$}& \multicolumn{2}{c}{$A_K>2$}  \\
%\cline{2-3}
\cmidrule(r){2-3} \cmidrule(r){4-5} \cmidrule(r){6-7} \cmidrule(r){8-9}
 & $\beta_\lambda$ & Number & $\beta_\lambda$ & Number & $\beta_\lambda$ & Number & $\beta_\lambda$ & Number\\
\midrule
$Z$     &-1.065$\pm$0.009 & 8348 &-1.165$\pm$0.012 & 5439 &-1.000$\pm$0.038 & 583 &-1.893$\pm$0.585 & 5 \\
$Y$     &-0.539$\pm$0.004 & 8710  &-0.590$\pm$0.007 & 5600 &-0.521$\pm$0.015 & 898 &-0.539$\pm$0.082 & 24 \\
$H$     & 0.660$\pm$0.004 & 8073 & 0.600$\pm$0.005 & 6103 & 0.591$\pm$0.008 & 1212 & 0.600$\pm$0.018 & 103 \\
$[3.6]$ & 1.144$\pm$0.006 & 9128 & 1.229$\pm$0.008 & 6100 & 1.207$\pm$0.014 & 1211 & 1.158$\pm$0.031 &  103\\
$[4.5]$ & 1.146$\pm$0.007 & 9132 & 1.276$\pm$0.010 & 6104 & 1.267$\pm$0.016 & 1213 & 1.201$\pm$0.031 & 103 \\
$[5.8]$ & 1.164$\pm$0.010 & 5730 & 1.347$\pm$0.012 & 4064 & 1.349$\pm$0.019 & 834  & 1.246$\pm$0.034 & 102 \\
$[8.0]$ & 1.168$\pm$0.014 & 3151  & 1.315$\pm$0.017 & 2225 & 1.260$\pm$0.027 & 473  & 1.211$\pm$0.035 & 77 \\
%\cmidrule{2-6}
\bottomrule
\end{tabularx}
%\textsuperscript{1}
%\noindent{\footnotesize{\textsuperscript{1} Tables may have a footer.}}
\end{adjustwidth}
\end{table}

\begin{table}[H] 
\centering
\caption{Results for the relative extinction $A_\lambda/A_K$ within various ranges of $A_K$. \label{tab:4}}
\newcolumntype{C}{>{\centering\arraybackslash}X}
%\begin{tabularx}{\textwidth}{CCCC}
\begin{tabularx}{13cm}{CCCCC}
\toprule
 &$A_K\leq0.5$ & $0.5<A_K\leq 1$ & $1<A_K\leq 2$& $A_K>2$\\
\midrule
$\beta_H$ & 0.660$\pm$0.004 &  0.600$\pm$0.005 & 0.591$\pm$0.008 & 0.600$\pm$0.018 \\
$\alpha$        & 2.777$\pm$0.064 & 1.848$\pm$0.075 & 1.713$\pm$0.119 & 1.848$\pm$0.270 \\
\midrule
$A_Z/A_K$       & 8.754$\pm$0.357 & 4.945$\pm$0.259 & 4.234$\pm$0.355 & 4.644$\pm$1.638 \\
$A_Y/A_K$       & 6.779$\pm$0.265 & 3.897$\pm$0.190 & 3.459$\pm$0.267 & 3.771$\pm$0.678 \\
$A_J/A_K$       & 4.755$\pm$0.172 & 2.822$\pm$0.119 & 2.617$\pm$0.175 & 2.822$\pm$0.430 \\
$A_H/A_K$       & 2.277$\pm$0.043 & 1.729$\pm$0.038 & 1.661$\pm$0.059 & 1.729$\pm$0.138 \\
$A_{[3.6]}/A_K$ & 0.458$\pm$0.033 & 0.583$\pm$0.031 & 0.665$\pm$0.043 & 0.623$\pm$0.088 \\
$A_{[4.5]}/A_K$ & 0.452$\pm$0.036 & 0.497$\pm$0.038 & 0.568$\pm$0.053 & 0.514$\pm$0.103 \\
$A_{[5.8]}/A_K$ & 0.384$\pm$0.047 & 0.368$\pm$0.047 & 0.436$\pm$0.068 & 0.364$\pm$0.123 \\
$A_{[8.0]}/A_K$ & 0.369$\pm$0.060 & 0.426$\pm$0.049 & 0.580$\pm$0.063 & 0.526$\pm$0.111 \\
%\cmidrule{2-6}
\bottomrule
\end{tabularx}
%\textsuperscript{1}
%\noindent{\footnotesize{\textsuperscript{1} Tables may have a footer.}}
\end{table}

%\unskip

For the sub-sample with $A_K\leq0.5$, we observed a greatly higher NIR color excess ratio of $E_{J-H}/E_{J-K}=0.66$ compared to the sub-samples with $A_K>0.5$ where the ratio was approximately $E_{J-H}/E_{J-K}\approx 0.6$. Consequently, the calculated values of $A_\lambda/A_K$ in the $ZYJH$ bands for this sub-sample exhibited a steeper slope compared to the other sub-samples. In general, the NIR $A_\lambda/A_K$ values exhibited a notable decrease with increasing $A_K$, except for the sub-sample with $A_K>2$, which displayed very large uncertainties. The sources with $A_K\leq0.5$ likely trace regions outside the dark clouds, while the sources with $A_K>0.5$ are more likely to be located within the interior of the dark clouds. This finding suggests that there is dust grain growth occurring with increasing extinction depth within the dark clouds. However, it is worth noting that the $A_\lambda/A_K$ values for the sub-sample with $A_K\leq0.5$ in the Spitzer/IRAC bands demonstrated a very flat extinction law, which is significantly flatter than the extinction law predicted by the WD01 $R_V$=3.1 model, and only slightly smaller than the WD01 $R_V$=5.5 model. This suggests the presence of a significant amount of large-sized dust grains in the outer regions of the dark cloud. One possible explanation for these observations is that the dust surrounding the dark cloud has undergone evolutionary growth, not limited to the interior of the cloud itself. Another hypothesis proposed by Ascenso et al.\cite{ascenso2013} is the influence of outflows generated by star formation activities within the dark cloud.

The variability of extinction laws has been a subject of significant controversy in previous studies. Equation \ref{equ:1} emphasizes the high sensitivity of the NIR color excess ratio to the effective wavelength chosen by photometric systems. Wang \& Jiang\cite{wang2014} pointed out that different photometric systems can lead to differences of more than 10\% when calculating $\alpha$ and $A_J/A_K$. Previous investigations have utilized diverse photometric filters, examined various interstellar environments, and explored a wide range of extinction depths. Wang \& Jiang\cite{wang2014} reported the universality of the NIR extinction law based on 2MASS photometry, finding $E_{J-H}/E_{J-K}=0.64$ and $\alpha=1.95$. Meingast et al.\cite{meingast2018} studied the extinction law in Orion using Visible and Infrared Survey Telescope for Astronomy (VISTA) data and obtained $E_{J-H}/E_{J-K}=0.636\pm0.002$, without discerning a significant trend in the NIR extinction law with respect to extinction depth. However, Kenyon et al.\cite{kenyon1998} discovered a trend of shallowing slope in the NIR extinction with increasing extinction, which was subsequently confirmed by studies conducted by Naoi et al.\cite{naoi2006,naoi2007}. In the MIR regime, Chapman et al.\cite{chapman2009} and McClure\cite{McClure2009} investigated star-forming regions and observed a flattening trend of $A_\lambda/A_K$ with increasing $A_K$. In contrast, Xue et al.\cite{xue2016} did not detect a noticeable trend in their measurements, which can be attributed to differences in the studied targets. Specifically, Xue et al.\cite{xue2016} primarily focused on dust grains in the diffuse ISM, whereas our study of the Ophiuchus cloud explores how a denser environment promotes dust growth.

\subsection{Spatial Variations across the Cloud}\label{subsec:varia}

\begin{figure}[H]
\centering
\includegraphics[width=9.5 cm]{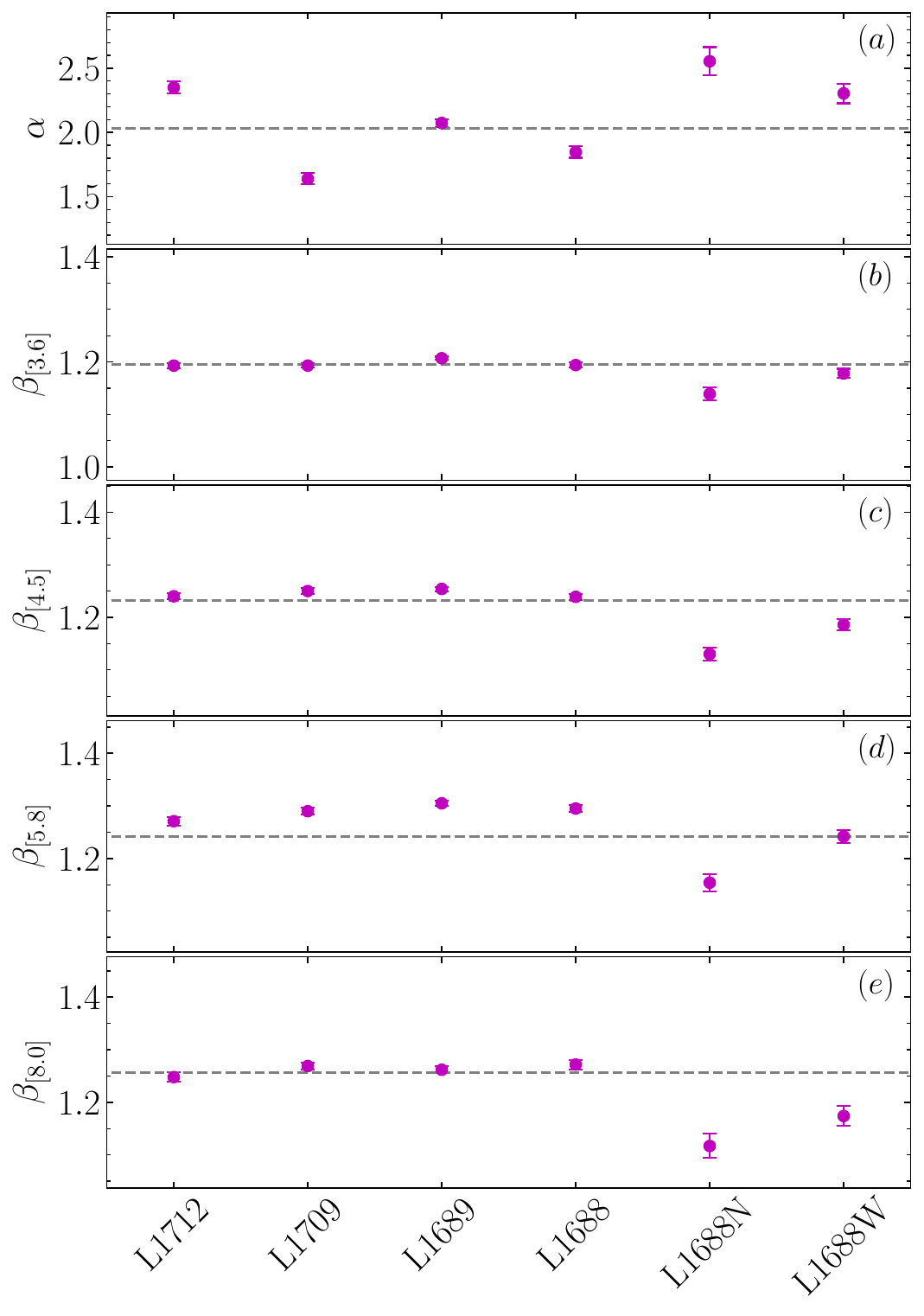}
\caption{The NIR extinction power-law index $\alpha$ and color excess ratios $\beta_\lambda=E_{J-\lambda}/E_{J-K}$ in the Spitzer/IRAC [3.6], [4.5], [5.8] and [8.0] bands for the sub-regions of Ophiuchus cloud as defined in Figure \ref{fig:ak-map}. The gray dashed lines represent the overall values of the parameters for the entire Ophiuchus cloud, as listed in Table \ref{tab:2}.\label{fig:beta-varia}}
\end{figure}

We now turn our attention to investigating the spatial variation of the extinction law within the Ophiuchus cloud. The Ophiuchus region consists of multiple dark clouds, namely L1712, L1689, L1709, and L1688, arranged in an east-to-west orientation (see Figure \ref{fig:ak-map}). Additionally, we selected two regions located to the north and west of L1688, referred to as L1688N and L1688W, respectively. For each of these sub-regions, we performed the fitting procedure and obtained the NIR extinction power-law index $\alpha$ by fitting the $J-H$ versus $J-K$ color-color diagram. The color excess ratios, $\beta_\lambda=E_{J-\lambda}/E_{J-K}$, were determined for the Spitzer/IRAC bands. Figure \ref{fig:beta-varia} displays the corresponding values of $\alpha$ and $\beta_\lambda=E_{J-\lambda}/E_{J-K}$ for these sub-regions.

Our findings reveal variations in the NIR extinction law within the same cloud, with differences exceeding 25\%. Specifically, regions L1712, L1688N, and L1688W, characterized by relatively low extinction, exhibit higher values of $\alpha$ compared to other regions. This trend is consistent with the results presented in Section \ref{subsec:depth}, where the sub-sample with lower-extinction demonstrates the higher value of $\alpha$ compared to other high-extinction samples. In a previous study by Naoi et al.\cite{naoi2007}, it was noted that L1712 exhibits a larger value of $E_{J-H}/E_{H-K}$ ($1.68\pm0.01$) compared to L1688 ($1.60\pm0.01$), based on observations using the SIRIUS infrared camera on the IRSF 1.4 m telescope at SAAO. Stead \& Hoare\cite{stead2009} investigated the NIR extinction slope using data from the UKIDSS Galactic Plane Survey and found a nearly constant $\alpha$ value of 2.14 across different sightlines. Maíz Apellániz et al.\cite{maiz2020} argue that different measurement methods may contribute to the observed variations in the NIR extinction, suggesting that the NIR extinction does not strictly follow a power law. In a study by Nogueras-Lara et al.\cite{Nogueras-Lara2019}, it was observed that the $\alpha$ value for the $JH$ band ($\alpha=2.43\pm0.10$) is larger than that for the $HK$ band ($\alpha=2.23\pm0.03$). This suggests that the extinction in the $K$ band tends to flatten, indicating that color excess relationships constructed based on the $JHK$ bands, such as $J-H$ vs. $J-K$ or $H-K$, still require further investigation.

Furthermore, we observed that the MIR color-excess ratio $\beta_\lambda=E_{J-\lambda}/E_{J-K}$ remains relatively unchanged in the four dark clouds, namely L1712, L1689, L1709, and L1688, which is consistent with the overall results obtained for the entire Ophiuchus molecular cloud in Section \ref{subsec:a-lambda}. However, the other two regions, L1688N and L1688W, exhibit relatively lower values of $\beta_\lambda$. These regions are located outside the dark clouds and have lower levels of extinction. This finding aligns with the conclusions presented in Section \ref{subsec:depth} and further supports the hypothesis of dust growth towards the inner regions of the dark clouds.

%\unskip

% Example of a page in landscape format (with table and table footnote).
%\startlandscape
%\begin{table}[H] %% Table in wide page
%\caption{This is a very wide table.\label{tab3}}
%	\begin{tabularx}{\textwidth}{CCCC}
%		\toprule
%		\textbf{Title 1}	& \textbf{Title 2}	& \textbf{Title 3}	& \textbf{Title 4}\\
%		\midrule
%		Entry 1		& Data			& Data			& This cell has some longer content that runs over two lines.\\
%		Entry 2		& Data			& Data			& Data\textsuperscript{1}\\
%		\bottomrule
%	\end{tabularx}
%	\begin{adjustwidth}{+\extralength}{0cm}
%		\noindent\footnotesize{\textsuperscript{1} This is a table footnote.}
%	\end{adjustwidth}
%\end{table}
%\finishlandscape

%%%%%%%%%%%%%%%%%%%%%%%%%%%%%%%%%%%%%%%%%%

%%%%%%%%%%%%%%%%%%%%%%%%%%%%%%%%%%%%%%%%%%
\section{Conclusions}\label{sec:sum}

In this study, we investigated the infrared extinction law of the nearby Ophiuchus molecular cloud using the NIR data from UKIDSS GCS and the MIR data from Spitzer c2d survey. To accurately analyze the observed colors, we carefully considered the non-linear effects arising from the response curves of the filters. We applied corrections based on the extinction law with WD01 $R_V=5.5$ and incorporated stellar template spectra. Employing the color-excess ratio method, we derived the relative extinction law $A_\lambda/A_K$ for the Ophiuchus molecular cloud, covering the UKIDSS $ZYJHK$ bands and the Spitzer/IRAC [3.6], [4.5], [5.8], and [8.0] bands. Additionally, we investigated the variations of the extinction law in relation to the depth of extinction and the spatial environment of the cloud. Our study yields the following main conclusions:

\begin{enumerate}
  \item We obtained the UKIDSS NIR color-excess ratios for the entire Ophiuchus cloud to be approximately $E_{J-H}/E_{J-K}\approx0.612$, corresponding to $A_J/A_K\approx3.123$ and $A_H/A_K\approx1.824$ when assuming a power-law NIR extinction law. The relative extinctions, compared to the $K$ band, were also calculated for other wavelength bands, and the measurement results are summarized in Table \ref{tab:2}. Additionally, we observed that the derived MIR extinction law in the $\sim3-8\,\mu$m range exhibits a flat behavior and closely resembles the WD01 model extinction law with $R_V$=5.5. In contrast, the NIR extinction law exhibited a steeper slope compared to the extinction law of WD01 $R_V$=5.5.
  \item We examined  the variation of the extinction law with increasing extinction depth by analyzing the average extinction laws of four sub-samples classified into different extinction bins: $A_K\leq0.5$, $0<A_K\leq1$, $1<A_K\leq2$, and $A_K>2$. Instead of assuming a uniform value of $A_J/A_K$, we calculated the specific $A_J/A_K$ for each sub-sample using $E_{J-H}/E_{J-K}$ and then derived $A_\lambda/A_K$ for other wavelength bands. Our analysis revealed a noticeable trend of a flatter extinction law from the NIR to MIR bands as the extinction increased. This observed trend is consistent with previous studies conducted by Naoi et al.\cite{naoi2006} and Chapman et al.\cite{chapman2009}.
  \item We also discussed the spatial variation of the extinction law within the Ophiuchus molecular cloud. Our analysis revealed no significant difference in the MIR extinction law among the four dark clouds, namely L1712, L1689, L1709, and L1688. However, we observed noticeable variations in the extinction law for the regions located outside the dark clouds, specifically L1688N and L1688W. These regions exhibited a lower color-excess ratio $E_{J-\lambda}/E_{J-K}$ in the Spitzer/IRAC bands. These findings provide further support for dust growth in the dense regions of the Ophiuchus cloud.
\end{enumerate}

%%%%%%%%%%%%%%%%%%%%%%%%%%%%%%%%%%%%%%%%%%
%\section{Patents}

%This section is not mandatory, but may be added if there %are patents resulting from the work reported in this manuscript.

%%%%%%%%%%%%%%%%%%%%%%%%%%%%%%%%%%%%%%%%%%
\vspace{6pt} 

%%%%%%%%%%%%%%%%%%%%%%%%%%%%%%%%%%%%%%%%%%
%% optional
%\supplementary{The following supporting information can be downloaded at:  \linksupplementary{s1}, Figure S1: title; Table S1: title; Video S1: title.}

% Only for the journal Methods and Protocols:
% If you wish to submit a video article, please do so with any other supplementary material.
% \supplementary{The following supporting information can be downloaded at: \linksupplementary{s1}, Figure S1: title; Table S1: title; Video S1: title. A supporting video article is available at doi: link.}

%%%%%%%%%%%%%%%%%%%%%%%%%%%%%%%%%%%%%%%%%%
\authorcontributions{Conceptualization, J.L. and X.C.; methodology, J.L. and X.C.; software, J.L.; validation, J.L.; formal analysis, J.L.; investigation, J.L. and X.C.; resources, J.L. and X.C.; data curation, J.L. and X.C.; writing---original draft preparation, J.L.; writing---review and editing, J.L. and X.C.; visualization, J.L. and X.C.; supervision, X.C.; project administration, J.L. and X.C.; funding acquisition, X.C. All authors have read and agreed to the published version of the manuscript.}

\funding{This research was funded by the National Key R\&D program of China (2022YFA1603102), the National Natural Science Foundation of China (11873002, 12011530065, 11590781). X.C. thanks to Guangdong Province Universities and Colleges Pearl River Scholar Funded Scheme (2019).}

\dataavailability{The data can be accessed from the WFCAM Science Archive (http://su rveys .roe.ac.uk/wsa) and the website of Spitzer Science Center (https://irsa.ipac.caltech.edu/data/S PITZER /C2D/).}

\acknowledgments{We are very grateful to the anonymous referees for their helpful comments and suggestions, which highly improved the paper. We thank Dr. He Zhao for his very helpful discussions.}

\conflictsofinterest{The authors declare no conflict of interest.} 

%%%%%%%%%%%%%%%%%%%%%%%%%%%%%%%%%%%%%%%%%%
%% Optional
%\sampleavailability{Samples of the compounds ... are available from the authors.}

%% Only for journal Encyclopedia
%\entrylink{The Link to this entry published on the encyclopedia platform.}

%\abbreviations{Abbreviations}{
%The following abbreviations are used in this manuscript:\\

%\noindent 
%\begin{tabular}{@{}ll}
%MDPI & Multidisciplinary Digital Publishing Institute\\
%DOAJ & Directory of open access journals\\
%TLA & Three letter acronym\\
%LD & Linear dichroism
%\end{tabular}
%}

%%%%%%%%%%%%%%%%%%%%%%%%%%%%%%%%%%%%%%%%%%
%% Optional

%%%%%%%%%%%%%%%%%%%%%%%%%%%%%%%%%%%%%%%%%%
\begin{adjustwidth}{-\extralength}{0cm}
%\printendnotes[custom] % Un-comment to print a list of endnotes

\reftitle{References}

% Please provide either the correct journal abbreviation (e.g. according to the “List of Title Word Abbreviations” http://www.issn.org/services/online-services/access-to-the-ltwa/) or the full name of the journal.
% Citations and References in Supplementary files are permitted provided that they also appear in the reference list here. 

%=====================================
% References, variant A: external bibliography
%=====================================
\bibliography{oph-ir-ext-law}

\begin{thebibliography}{999}

\bibitem[{Draine}(2003)]{draine2003}
{Draine}, B.T.
\newblock {Interstellar Dust Grains}.
\newblock {\em \araa} {\bf 2003}, {\em 41},~241--289,
  \href{http://xxx.lanl.gov/abs/astro-ph/0304489}{{\normalfont
  [arXiv:astro-ph/astro-ph/0304489]}}.
\newblock {\url{https://doi.org/10.1146/annurev.astro.41.011802.094840}}.

\bibitem[{Zubko} \em{et~al.}(2004){Zubko}, {Dwek}, and {Arendt}]{Zubko2004}
{Zubko}, V.; {Dwek}, E.; {Arendt}, R.G.
\newblock {Interstellar Dust Models Consistent with Extinction, Emission, and
  Abundance Constraints}.
\newblock {\em \apjs} {\bf 2004}, {\em 152},~211--249,
  \href{http://xxx.lanl.gov/abs/astro-ph/0312641}{{\normalfont
  [arXiv:astro-ph/astro-ph/0312641]}}.
\newblock {\url{https://doi.org/10.1086/382351}}.

\bibitem[{Compi{\`e}gne} \em{et~al.}(2011){Compi{\`e}gne}, {Verstraete},
  {Jones}, {Bernard}, {Boulanger}, {Flagey}, {Le Bourlot}, {Paradis}, and
  {Ysard}]{Compiegne2011}
{Compi{\`e}gne}, M.; {Verstraete}, L.; {Jones}, A.; {Bernard}, J.P.;
  {Boulanger}, F.; {Flagey}, N.; {Le Bourlot}, J.; {Paradis}, D.; {Ysard}, N.
\newblock {The global dust SED: tracing the nature and evolution of dust with
  DustEM}.
\newblock {\em \aap} {\bf 2011}, {\em 525},~A103,
  \href{http://xxx.lanl.gov/abs/1010.2769}{{\normalfont
  [arXiv:astro-ph.GA/1010.2769]}}.
\newblock {\url{https://doi.org/10.1051/0004-6361/201015292}}.

\bibitem[{Hensley} and {Draine}(2021)]{Hensley2021}
{Hensley}, B.S.; {Draine}, B.T.
\newblock {Observational Constraints on the Physical Properties of Interstellar
  Dust in the Post-Planck Era}.
\newblock {\em \apj} {\bf 2021}, {\em 906},~73,
  \href{http://xxx.lanl.gov/abs/2009.00018}{{\normalfont
  [arXiv:astro-ph.GA/2009.00018]}}.
\newblock {\url{https://doi.org/10.3847/1538-4357/abc8f1}}.

\bibitem[{Whittet} \em{et~al.}(1998){Whittet}, {Gerakines}, {Tielens},
  {Adamson}, {Boogert}, {Chiar}, {de Graauw}, {Ehrenfreund}, {Prusti},
  {Schutte}, {Vandenbussche}, and {van Dishoeck}]{Whittet1998}
{Whittet}, D.C.B.; {Gerakines}, P.A.; {Tielens}, A.G.G.M.; {Adamson}, A.J.;
  {Boogert}, A.C.A.; {Chiar}, J.E.; {de Graauw}, T.; {Ehrenfreund}, P.;
  {Prusti}, T.; {Schutte}, W.A.;  et~al.
\newblock {Detection of Abundant CO$_{2}$ Ice in the Quiescent Dark Cloud
  Medium toward Elias 16}.
\newblock {\em \apjl} {\bf 1998}, {\em 498},~L159--L163.
\newblock {\url{https://doi.org/10.1086/311318}}.

\bibitem[{Ossenkopf} and {Henning}(1994)]{ossenkopf1994}
{Ossenkopf}, V.; {Henning}, T.
\newblock {Dust opacities for protostellar cores.}
\newblock {\em \aap} {\bf 1994}, {\em 291},~943--959.

\bibitem[{Ormel} \em{et~al.}(2011){Ormel}, {Min}, {Tielens}, {Dominik}, and
  {Paszun}]{ormel2011}
{Ormel}, C.W.; {Min}, M.; {Tielens}, A.G.G.M.; {Dominik}, C.; {Paszun}, D.
\newblock {Dust coagulation and fragmentation in molecular clouds. II. The
  opacity of the dust aggregate size distribution}.
\newblock {\em \aap} {\bf 2011}, {\em 532},~A43,
  \href{http://xxx.lanl.gov/abs/1106.3265}{{\normalfont
  [arXiv:astro-ph.SR/1106.3265]}}.
\newblock {\url{https://doi.org/10.1051/0004-6361/201117058}}.

\bibitem[{Lada} \em{et~al.}(1994){Lada}, {Lada}, {Clemens}, and
  {Bally}]{Lada1994}
{Lada}, C.J.; {Lada}, E.A.; {Clemens}, D.P.; {Bally}, J.
\newblock {Dust Extinction and Molecular Gas in the Dark Cloud IC 5146}.
\newblock {\em \apj} {\bf 1994}, {\em 429},~694.
\newblock {\url{https://doi.org/10.1086/174354}}.

\bibitem[{Lombardi} and {Alves}(2001)]{Lombardi2001}
{Lombardi}, M.; {Alves}, J.
\newblock {Mapping the interstellar dust with near-infrared observations: An
  optimized multi-band technique}.
\newblock {\em \aap} {\bf 2001}, {\em 377},~1023--1034,
  \href{http://xxx.lanl.gov/abs/astro-ph/0109135}{{\normalfont
  [arXiv:astro-ph/astro-ph/0109135]}}.
\newblock {\url{https://doi.org/10.1051/0004-6361:20011099}}.

\bibitem[{Alves} \em{et~al.}(2001){Alves}, {Lada}, and {Lada}]{Alves2001}
{Alves}, J.F.; {Lada}, C.J.; {Lada}, E.A.
\newblock {Internal structure of a cold dark molecular cloud inferred from the
  extinction of background starlight}.
\newblock {\em Nature} {\bf 2001}, {\em 409},~159--161.

\bibitem[{Fitzpatrick}(1999)]{Fitzpatrick1999}
{Fitzpatrick}, E.L.
\newblock {Correcting for the Effects of Interstellar Extinction}.
\newblock {\em \pasp} {\bf 1999}, {\em 111},~63--75,
  \href{http://xxx.lanl.gov/abs/astro-ph/9809387}{{\normalfont
  [arXiv:astro-ph/astro-ph/9809387]}}.
\newblock {\url{https://doi.org/10.1086/316293}}.

\bibitem[{Cardelli} \em{et~al.}(1989){Cardelli}, {Clayton}, and
  {Mathis}]{cardelli1989}
{Cardelli}, J.A.; {Clayton}, G.C.; {Mathis}, J.S.
\newblock {The relationship between infrared, optical, and ultraviolet
  extinction}.
\newblock {\em \apj} {\bf 1989}, {\em 345},~245--256.
\newblock {\url{https://doi.org/10.1086/167900}}.

\bibitem[{Wang} and {Jiang}(2014)]{wang2014}
{Wang}, S.; {Jiang}, B.W.
\newblock {Universality of the Near-infrared Extinction Law Based on the APOGEE
  Survey}.
\newblock {\em \apjl} {\bf 2014}, {\em 788},~L12.
\newblock {\url{https://doi.org/10.1088/2041-8205/788/1/L12}}.

\bibitem[{Schultheis} \em{et~al.}(2015){Schultheis}, {Kordopatis},
  {Recio-Blanco}, {de Laverny}, {Hill}, {Gilmore}, {Alfaro}, {Costado},
  {Bensby}, {Damiani}, {Feltzing}, {Flaccomio}, {Lardo}, {Jofre}, {Prisinzano},
  {Zaggia}, {Jimenez-Esteban}, {Morbidelli}, {Lanzafame}, {Hourihane},
  {Worley}, and {Francois}]{Schultheis2015}
{Schultheis}, M.; {Kordopatis}, G.; {Recio-Blanco}, A.; {de Laverny}, P.;
  {Hill}, V.; {Gilmore}, G.; {Alfaro}, E.J.; {Costado}, M.T.; {Bensby}, T.;
  {Damiani}, F.;  et~al.
\newblock {The Gaia-ESO Survey: Tracing interstellar extinction}.
\newblock {\em \aap} {\bf 2015}, {\em 577},~A77,
  \href{http://xxx.lanl.gov/abs/1502.03223}{{\normalfont
  [arXiv:astro-ph.GA/1502.03223]}}.
\newblock {\url{https://doi.org/10.1051/0004-6361/201425333}}.

\bibitem[{Matsunaga} \em{et~al.}(2018){Matsunaga}, {Bono}, {Chen}, {de Grijs},
  {Inno}, and {Nishiyama}]{matsunaga2018}
{Matsunaga}, N.; {Bono}, G.; {Chen}, X.; {de Grijs}, R.; {Inno}, L.;
  {Nishiyama}, S.
\newblock {Impact of Distance Determinations on Galactic Structure. I. Young
  and Intermediate-Age Tracers}.
\newblock {\em Space Science Reviews} {\bf 2018}, {\em 214},~74,
  \href{http://xxx.lanl.gov/abs/1804.04931}{{\normalfont
  [arXiv:astro-ph.SR/1804.04931]}}.
\newblock {\url{https://doi.org/10.1007/s11214-018-0506-5}}.

\bibitem[{Rom{\'a}n-Z{\'u}{\~n}iga}
  \em{et~al.}(2007){Rom{\'a}n-Z{\'u}{\~n}iga}, {Lada}, {Muench}, and
  {Alves}]{roman2007ApJ}
{Rom{\'a}n-Z{\'u}{\~n}iga}, C.G.; {Lada}, C.J.; {Muench}, A.; {Alves}, J.F.
\newblock {The Infrared Extinction Law at Extreme Depth in a Dark Cloud Core}.
\newblock {\em \apj} {\bf 2007}, {\em 664},~357--362,
  \href{http://xxx.lanl.gov/abs/0704.3203}{{\normalfont
  [arXiv:astro-ph/0704.3203]}}.
\newblock {\url{https://doi.org/10.1086/518928}}.

\bibitem[{Gao} \em{et~al.}(2009){Gao}, {Jiang}, and {Li}]{Gao2009}
{Gao}, J.; {Jiang}, B.W.; {Li}, A.
\newblock {Mid-Infrared Extinction and its Variation with Galactic Longitude}.
\newblock {\em \apj} {\bf 2009}, {\em 707},~89--102,
  \href{http://xxx.lanl.gov/abs/0910.3037}{{\normalfont
  [arXiv:astro-ph.GA/0910.3037]}}.
\newblock {\url{https://doi.org/10.1088/0004-637X/707/1/89}}.

\bibitem[{Weingartner} and {Draine}(2001)]{weingartner2001}
{Weingartner}, J.C.; {Draine}, B.T.
\newblock {Dust Grain-Size Distributions and Extinction in the Milky Way, Large
  Magellanic Cloud, and Small Magellanic Cloud}.
\newblock {\em \apj} {\bf 2001}, {\em 548},~296--309,
  \href{http://xxx.lanl.gov/abs/astro-ph/0008146}{{\normalfont
  [arXiv:astro-ph/astro-ph/0008146]}}.
\newblock {\url{https://doi.org/10.1086/318651}}.

\bibitem[{Cambr{\'e}sy} \em{et~al.}(2011){Cambr{\'e}sy}, {Rho}, {Marshall}, and
  {Reach}]{cambresy2011}
{Cambr{\'e}sy}, L.; {Rho}, J.; {Marshall}, D.J.; {Reach}, W.T.
\newblock {Variation of the extinction law in the Trifid nebula}.
\newblock {\em \aap} {\bf 2011}, {\em 527},~A141,
  \href{http://xxx.lanl.gov/abs/1101.1089}{{\normalfont
  [arXiv:astro-ph.GA/1101.1089]}}.
\newblock {\url{https://doi.org/10.1051/0004-6361/201015863}}.

\bibitem[{Wang} \em{et~al.}(2015{\natexlab{a}}){Wang}, {Li}, and
  {Jiang}]{wang2015a}
{Wang}, S.; {Li}, A.; {Jiang}, B.W.
\newblock {Very Large Interstellar Grains as Evidenced by the Mid-infrared
  Extinction}.
\newblock {\em \apj} {\bf 2015}, {\em 811},~38,
  \href{http://xxx.lanl.gov/abs/1508.03403}{{\normalfont
  [arXiv:astro-ph.GA/1508.03403]}}.
\newblock {\url{https://doi.org/10.1088/0004-637X/811/1/38}}.

\bibitem[{Wang} \em{et~al.}(2015{\natexlab{b}}){Wang}, {Li}, and
  {Jiang}]{wang2015b}
{Wang}, S.; {Li}, A.; {Jiang}, B.W.
\newblock {The interstellar oxygen crisis, or where have all the oxygen atoms
  gone?}
\newblock {\em \mnras} {\bf 2015}, {\em 454},~569--575,
  \href{http://xxx.lanl.gov/abs/1508.03404}{{\normalfont
  [arXiv:astro-ph.GA/1508.03404]}}.
\newblock {\url{https://doi.org/10.1093/mnras/stv1900}}.

\bibitem[{Ascenso} \em{et~al.}(2013){Ascenso}, {Lada}, {Alves},
  {Rom{\'a}n-Z{\'u}{\~n}iga}, and {Lombardi}]{ascenso2013}
{Ascenso}, J.; {Lada}, C.J.; {Alves}, J.; {Rom{\'a}n-Z{\'u}{\~n}iga}, C.G.;
  {Lombardi}, M.
\newblock {The mid-infrared extinction law in the darkest cores of the Pipe
  Nebula}.
\newblock {\em \aap} {\bf 2013}, {\em 549},~A135,
  \href{http://xxx.lanl.gov/abs/1211.6556}{{\normalfont
  [arXiv:astro-ph.GA/1211.6556]}}.
\newblock {\url{https://doi.org/10.1051/0004-6361/201220658}}.

\bibitem[{Mamajek}(2008)]{Mamajek2008}
{Mamajek}, E.E.
\newblock {On the distance to the Ophiuchus star-forming region}.
\newblock {\em Astronomische Nachrichten} {\bf 2008}, {\em 329},~10,
  \href{http://xxx.lanl.gov/abs/0709.0505}{{\normalfont
  [arXiv:astro-ph/0709.0505]}}.
\newblock {\url{https://doi.org/10.1002/asna.200710827}}.

\bibitem[{Casewell} and {Hambly}(2013)]{Casewell2013}
{Casewell}, S.; {Hambly}, N.
\newblock {The UKIDSS Galactic Clusters Survey}.
\newblock In Proceedings of the Thirty Years of Astronomical Discovery with
  UKIRT,  2013, Vol.~37, {\em Astrophysics and Space Science Proceedings}, p.
  291.
\newblock {\url{https://doi.org/10.1007/978-94-007-7432-2_27}}.

\bibitem[{Lawrence} \em{et~al.}(2007){Lawrence}, {Warren}, {Almaini}, {Edge},
  {Hambly}, {Jameson}, {Lucas}, {Casali}, {Adamson}, {Dye}, {Emerson},
  {Foucaud}, {Hewett}, {Hirst}, {Hodgkin}, {Irwin}, {Lodieu}, {McMahon},
  {Simpson}, {Smail}, {Mortlock}, and {Folger}]{Lawrence2007}
{Lawrence}, A.; {Warren}, S.J.; {Almaini}, O.; {Edge}, A.C.; {Hambly}, N.C.;
  {Jameson}, R.F.; {Lucas}, P.; {Casali}, M.; {Adamson}, A.; {Dye}, S.;  et~al.
\newblock {The UKIRT Infrared Deep Sky Survey (UKIDSS)}.
\newblock {\em \mnras} {\bf 2007}, {\em 379},~1599--1617,
  \href{http://xxx.lanl.gov/abs/astro-ph/0604426}{{\normalfont
  [arXiv:astro-ph/astro-ph/0604426]}}.
\newblock {\url{https://doi.org/10.1111/j.1365-2966.2007.12040.x}}.

\bibitem[{Hewett} \em{et~al.}(2006){Hewett}, {Warren}, {Leggett}, and
  {Hodgkin}]{Hewett2006}
{Hewett}, P.C.; {Warren}, S.J.; {Leggett}, S.K.; {Hodgkin}, S.T.
\newblock {The UKIRT Infrared Deep Sky Survey ZY JHK photometric system:
  passbands and synthetic colours}.
\newblock {\em \mnras} {\bf 2006}, {\em 367},~454--468,
  \href{http://xxx.lanl.gov/abs/astro-ph/0601592}{{\normalfont
  [arXiv:astro-ph/astro-ph/0601592]}}.
\newblock {\url{https://doi.org/10.1111/j.1365-2966.2005.09969.x}}.

\bibitem[{Hodgkin} \em{et~al.}(2009){Hodgkin}, {Irwin}, {Hewett}, and
  {Warren}]{Hodgkin2009}
{Hodgkin}, S.T.; {Irwin}, M.J.; {Hewett}, P.C.; {Warren}, S.J.
\newblock {The UKIRT wide field camera ZYJHK photometric system: calibration
  from 2MASS}.
\newblock {\em \mnras} {\bf 2009}, {\em 394},~675--692,
  \href{http://xxx.lanl.gov/abs/0812.3081}{{\normalfont
  [arXiv:astro-ph/0812.3081]}}.
\newblock {\url{https://doi.org/10.1111/j.1365-2966.2008.14387.x}}.

\bibitem[{Evans} \em{et~al.}(2003){Evans}, {Allen}, {Blake}, {Boogert},
  {Bourke}, {Harvey}, {Kessler}, {Koerner}, {Lee}, {Mundy}, {Myers}, {Padgett},
  {Pontoppidan}, {Sargent}, {Stapelfeldt}, {van Dishoeck}, {Young}, and
  {Young}]{evans2003}
{Evans}, Neal~J., I.; {Allen}, L.E.; {Blake}, G.A.; {Boogert}, A.C.A.;
  {Bourke}, T.; {Harvey}, P.M.; {Kessler}, J.E.; {Koerner}, D.W.; {Lee}, C.W.;
  {Mundy}, L.G.;  et~al.
\newblock {From Molecular Cores to Planet-forming Disks: An SIRTF Legacy
  Program}.
\newblock {\em \pasp} {\bf 2003}, {\em 115},~965--980,
  \href{http://xxx.lanl.gov/abs/astro-ph/0305127}{{\normalfont
  [arXiv:astro-ph/astro-ph/0305127]}}.
\newblock {\url{https://doi.org/10.1086/376697}}.

\bibitem[{Juvela} and {Montillaud}(2016)]{juvela2016}
{Juvela}, M.; {Montillaud}, J.
\newblock {Allsky NICER and NICEST extinction maps based on the 2MASS
  near-infrared survey}.
\newblock {\em \aap} {\bf 2016}, {\em 585},~A38,
  \href{http://xxx.lanl.gov/abs/1511.00883}{{\normalfont
  [arXiv:astro-ph.GA/1511.00883]}}.
\newblock {\url{https://doi.org/10.1051/0004-6361/201425112}}.

\bibitem[{Lombardi} \em{et~al.}(2008){Lombardi}, {Lada}, and
  {Alves}]{Lombardi2008}
{Lombardi}, M.; {Lada}, C.J.; {Alves}, J.
\newblock {2MASS wide field extinction maps. II. The Ophiuchus and the Lupus
  cloud complexes}.
\newblock {\em \aap} {\bf 2008}, {\em 489},~143--156,
  \href{http://xxx.lanl.gov/abs/0809.3740}{{\normalfont
  [arXiv:astro-ph/0809.3740]}}.
\newblock {\url{https://doi.org/10.1051/0004-6361:200810070}}.

\bibitem[{Taylor}(2005)]{Taylor2005}
{Taylor}, M.B.
\newblock {TOPCAT \& STIL: Starlink Table/VOTable Processing Software}.
\newblock In Proceedings of the Astronomical Data Analysis Software and Systems
  XIV; {Shopbell}, P.; {Britton}, M.; {Ebert}, R., Eds.,  2005, Vol. 347, {\em
  Astronomical Society of the Pacific Conference Series}, p.~29.

\bibitem[{Grasser} \em{et~al.}(2021){Grasser}, {Ratzenb{\"o}ck}, {Alves},
  {Gro{\ss}schedl}, {Meingast}, {Zucker}, {Hacar}, {Lada}, {Goodman},
  {Lombardi}, {Forbes}, {Bomze}, and {M{\"o}ller}]{grasser2021}
{Grasser}, N.; {Ratzenb{\"o}ck}, S.; {Alves}, J.; {Gro{\ss}schedl}, J.;
  {Meingast}, S.; {Zucker}, C.; {Hacar}, A.; {Lada}, C.; {Goodman}, A.;
  {Lombardi}, M.;  et~al.
\newblock {The {\ensuremath{\rho}} Ophiuchi region revisited with Gaia EDR3.
  Two young populations, new members, and old impostors}.
\newblock {\em \aap} {\bf 2021}, {\em 652},~A2.
\newblock {\url{https://doi.org/10.1051/0004-6361/202140438}}.

\bibitem[{Indebetouw} \em{et~al.}(2005){Indebetouw}, {Mathis}, {Babler},
  {Meade}, {Watson}, {Whitney}, {Wolff}, {Wolfire}, {Cohen}, {Bania},
  {Benjamin}, {Clemens}, {Dickey}, {Jackson}, {Kobulnicky}, {Marston},
  {Mercer}, {Stauffer}, {Stolovy}, and {Churchwell}]{indebetouw2005}
{Indebetouw}, R.; {Mathis}, J.S.; {Babler}, B.L.; {Meade}, M.R.; {Watson}, C.;
  {Whitney}, B.A.; {Wolff}, M.J.; {Wolfire}, M.G.; {Cohen}, M.; {Bania}, T.M.;
  et~al.
\newblock {The Wavelength Dependence of Interstellar Extinction from 1.25 to
  8.0 {\ensuremath{\mu}}m Using GLIMPSE Data}.
\newblock {\em \apj} {\bf 2005}, {\em 619},~931--938,
  \href{http://xxx.lanl.gov/abs/astro-ph/0406403}{{\normalfont
  [arXiv:astro-ph/astro-ph/0406403]}}.
\newblock {\url{https://doi.org/10.1086/426679}}.

\bibitem[{Stead} and {Hoare}(2009)]{stead2009}
{Stead}, J.J.; {Hoare}, M.G.
\newblock {The slope of the near-infrared extinction law}.
\newblock {\em \mnras} {\bf 2009}, {\em 400},~731--742,
  \href{http://xxx.lanl.gov/abs/0908.1601}{{\normalfont
  [arXiv:astro-ph.GA/0908.1601]}}.
\newblock {\url{https://doi.org/10.1111/j.1365-2966.2009.15530.x}}.

\bibitem[{Wang} and {Chen}(2019)]{wang2019}
{Wang}, S.; {Chen}, X.
\newblock {The Optical to Mid-infrared Extinction Law Based on the APOGEE, Gaia
  DR2, Pan-STARRS1, SDSS, APASS, 2MASS, and WISE Surveys}.
\newblock {\em \apj} {\bf 2019}, {\em 877},~116,
  \href{http://xxx.lanl.gov/abs/1904.04575}{{\normalfont
  [arXiv:astro-ph.GA/1904.04575]}}.
\newblock {\url{https://doi.org/10.3847/1538-4357/ab1c61}}.

\bibitem[{Meingast} \em{et~al.}(2018){Meingast}, {Alves}, and
  {Lombardi}]{meingast2018}
{Meingast}, S.; {Alves}, J.; {Lombardi}, M.
\newblock {VISION - Vienna Survey in Orion. II. Infrared extinction in Orion
  A}.
\newblock {\em \aap} {\bf 2018}, {\em 614},~A65,
  \href{http://xxx.lanl.gov/abs/1803.01004}{{\normalfont
  [arXiv:astro-ph.GA/1803.01004]}}.
\newblock {\url{https://doi.org/10.1051/0004-6361/201731396}}.

\bibitem[{Castelli} and {Kurucz}(2003)]{castelli2003}
{Castelli}, F.; {Kurucz}, R.L.
\newblock {New Grids of ATLAS9 Model Atmospheres}.
\newblock In Proceedings of the Modelling of Stellar Atmospheres; {Piskunov},
  N.; {Weiss}, W.W.; {Gray}, D.F., Eds.,  2003, Vol. 210, p. A20,
  \href{http://xxx.lanl.gov/abs/astro-ph/0405087}{{\normalfont
  [arXiv:astro-ph/astro-ph/0405087]}}.
\newblock {\url{https://doi.org/10.48550/arXiv.astro-ph/0405087}}.

\bibitem[{Sanders} \em{et~al.}(2022){Sanders}, {Smith},
  {Gonz{\'a}lez-Fern{\'a}ndez}, {Lucas}, and {Minniti}]{sanders2022}
{Sanders}, J.L.; {Smith}, L.; {Gonz{\'a}lez-Fern{\'a}ndez}, C.; {Lucas}, P.;
  {Minniti}, D.
\newblock {The extinction law in the inner 3 {\texttimes} 3 deg$^{2}$ of the
  Milky Way and the red clump absolute magnitude in the inner bar-bulge}.
\newblock {\em \mnras} {\bf 2022}, {\em 514},~2407--2424,
  \href{http://xxx.lanl.gov/abs/2205.10378}{{\normalfont
  [arXiv:astro-ph.GA/2205.10378]}}.
\newblock {\url{https://doi.org/10.1093/mnras/stac1367}}.

\bibitem[{Harris} \em{et~al.}(1978){Harris}, {Woolf}, and {Rieke}]{Harris1978}
{Harris}, D.H.; {Woolf}, N.J.; {Rieke}, G.H.
\newblock {Ice mantles and abnormal extinction in the Rho Ophiuchi cloud.}
\newblock {\em \apj} {\bf 1978}, {\em 226},~829--838.
\newblock {\url{https://doi.org/10.1086/156663}}.

\bibitem[{Kenyon} \em{et~al.}(1998){Kenyon}, {Lada}, and {Barsony}]{kenyon1998}
{Kenyon}, S.J.; {Lada}, E.A.; {Barsony}, M.
\newblock {The Near-Infrared Extinction Law and Limits on the Pre-Main-Sequence
  Population of the rho Ophiuchi Dark Cloud}.
\newblock {\em \aj} {\bf 1998}, {\em 115},~252--262.
\newblock {\url{https://doi.org/10.1086/300188}}.

\bibitem[{Naoi} \em{et~al.}(2006){Naoi}, {Tamura}, {Nakajima}, {Nagata},
  {Suto}, {Murakawa}, {Kandori}, {Sasaki}, {Baba}, {Kato}, {Kurita},
  {Nagashima}, {Nagayama}, {Nakaya}, {Nishiyama}, {Oasa}, {Sato}, and
  {Sugitani}]{naoi2006}
{Naoi}, T.; {Tamura}, M.; {Nakajima}, Y.; {Nagata}, T.; {Suto}, H.; {Murakawa},
  K.; {Kandori}, R.; {Sasaki}, S.; {Baba}, D.; {Kato}, D.;  et~al.
\newblock {Near-Infrared Extinction Law in the {\ensuremath{\rho}} Ophiuchi and
  Chamaeleon Dark Clouds}.
\newblock {\em \apj} {\bf 2006}, {\em 640},~373--382.
\newblock {\url{https://doi.org/10.1086/500112}}.

\bibitem[{Rieke} and {Lebofsky}(1985)]{rieke1985}
{Rieke}, G.H.; {Lebofsky}, M.J.
\newblock {The interstellar extinction law from 1 to 13 microns.}
\newblock {\em \apj} {\bf 1985}, {\em 288},~618--621.
\newblock {\url{https://doi.org/10.1086/162827}}.

\bibitem[{Nishiyama} \em{et~al.}(2009){Nishiyama}, {Tamura}, {Hatano}, {Kato},
  {Tanab{\'e}}, {Sugitani}, and {Nagata}]{nishiyama2009}
{Nishiyama}, S.; {Tamura}, M.; {Hatano}, H.; {Kato}, D.; {Tanab{\'e}}, T.;
  {Sugitani}, K.; {Nagata}, T.
\newblock {Interstellar Extinction Law Toward the Galactic Center III: J, H,
  K$_{S}$ Bands in the 2MASS and the MKO Systems, and 3.6, 4.5, 5.8, 8.0
  {\ensuremath{\mu}}m in the Spitzer/IRAC System}.
\newblock {\em \apj} {\bf 2009}, {\em 696},~1407--1417,
  \href{http://xxx.lanl.gov/abs/0902.3095}{{\normalfont
  [arXiv:astro-ph.GA/0902.3095]}}.
\newblock {\url{https://doi.org/10.1088/0004-637X/696/2/1407}}.

\bibitem[{Fritz} \em{et~al.}(2011){Fritz}, {Gillessen}, {Dodds-Eden}, {Lutz},
  {Genzel}, {Raab}, {Ott}, {Pfuhl}, {Eisenhauer}, and {Yusef-Zadeh}]{fritz2011}
{Fritz}, T.K.; {Gillessen}, S.; {Dodds-Eden}, K.; {Lutz}, D.; {Genzel}, R.;
  {Raab}, W.; {Ott}, T.; {Pfuhl}, O.; {Eisenhauer}, F.; {Yusef-Zadeh}, F.
\newblock {Line Derived Infrared Extinction toward the Galactic Center}.
\newblock {\em \apj} {\bf 2011}, {\em 737},~73,
  \href{http://xxx.lanl.gov/abs/1105.2822}{{\normalfont
  [arXiv:astro-ph.GA/1105.2822]}}.
\newblock {\url{https://doi.org/10.1088/0004-637X/737/2/73}}.

\bibitem[{Gordon} \em{et~al.}(2021){Gordon}, {Misselt}, {Bouwman}, {Clayton},
  {Decleir}, {Hines}, {Pendleton}, {Rieke}, {Smith}, and {Whittet}]{Gordon2021}
{Gordon}, K.D.; {Misselt}, K.A.; {Bouwman}, J.; {Clayton}, G.C.; {Decleir}, M.;
  {Hines}, D.C.; {Pendleton}, Y.; {Rieke}, G.; {Smith}, J.D.T.; {Whittet},
  D.C.B.
\newblock {Milky Way Mid-Infrared Spitzer Spectroscopic Extinction Curves:
  Continuum and Silicate Features}.
\newblock {\em \apj} {\bf 2021}, {\em 916},~33,
  \href{http://xxx.lanl.gov/abs/2105.05087}{{\normalfont
  [arXiv:astro-ph.GA/2105.05087]}}.
\newblock {\url{https://doi.org/10.3847/1538-4357/ac00b7}}.

\bibitem[{Chapman} \em{et~al.}(2009){Chapman}, {Mundy}, {Lai}, and
  {Evans}]{chapman2009}
{Chapman}, N.L.; {Mundy}, L.G.; {Lai}, S.P.; {Evans}, Neal~J., I.
\newblock {The Mid-Infrared Extinction Law in the Ophiuchus, Perseus, and
  Serpens Molecular Clouds}.
\newblock {\em \apj} {\bf 2009}, {\em 690},~496--511,
  \href{http://xxx.lanl.gov/abs/0809.1106}{{\normalfont
  [arXiv:astro-ph/0809.1106]}}.
\newblock {\url{https://doi.org/10.1088/0004-637X/690/1/496}}.

\bibitem[{Flaherty} \em{et~al.}(2007){Flaherty}, {Pipher}, {Megeath},
  {Winston}, {Gutermuth}, {Muzerolle}, {Allen}, and {Fazio}]{flaherty}
{Flaherty}, K.M.; {Pipher}, J.L.; {Megeath}, S.T.; {Winston}, E.M.;
  {Gutermuth}, R.A.; {Muzerolle}, J.; {Allen}, L.E.; {Fazio}, G.G.
\newblock {Infrared Extinction toward Nearby Star-forming Regions}.
\newblock {\em \apj} {\bf 2007}, {\em 663},~1069--1082,
  \href{http://xxx.lanl.gov/abs/astro-ph/0703777}{{\normalfont
  [arXiv:astro-ph/astro-ph/0703777]}}.
\newblock {\url{https://doi.org/10.1086/518411}}.

\bibitem[{Naoi} \em{et~al.}(2007){Naoi}, {Tamura}, {Nagata}, {Nakajima},
  {Suto}, {Murakawa}, {Kandori}, {Sasaki}, {Nishiyama}, {Oasa}, and
  {Sugitani}]{naoi2007}
{Naoi}, T.; {Tamura}, M.; {Nagata}, T.; {Nakajima}, Y.; {Suto}, H.; {Murakawa},
  K.; {Kandori}, R.; {Sasaki}, S.; {Nishiyama}, S.; {Oasa}, Y.;  et~al.
\newblock {Near-Infrared Extinction in the Coalsack Globule 2}.
\newblock {\em \apj} {\bf 2007}, {\em 658},~1114--1118,
  \href{http://xxx.lanl.gov/abs/astro-ph/0612620}{{\normalfont
  [arXiv:astro-ph/astro-ph/0612620]}}.
\newblock {\url{https://doi.org/10.1086/512030}}.

\bibitem[{McClure}(2009)]{McClure2009}
{McClure}, M.
\newblock {Observational 5-20 {\ensuremath{\mu}}m Interstellar Extinction
  Curves Toward Star-Forming Regions Derived From Spitzer IRS Spectra}.
\newblock {\em \apjl} {\bf 2009}, {\em 693},~L81--L85,
  \href{http://xxx.lanl.gov/abs/0810.4561}{{\normalfont
  [arXiv:astro-ph/0810.4561]}}.
\newblock {\url{https://doi.org/10.1088/0004-637X/693/2/L81}}.

\bibitem[{Xue} \em{et~al.}(2016){Xue}, {Jiang}, {Gao}, {Liu}, {Wang}, and
  {Li}]{xue2016}
{Xue}, M.; {Jiang}, B.W.; {Gao}, J.; {Liu}, J.; {Wang}, S.; {Li}, A.
\newblock {A Precise Determination of the Mid-infrared Interstellar Extinction
  Law Based on the APOGEE Spectroscopic Survey}.
\newblock {\em \apjs} {\bf 2016}, {\em 224},~23,
  \href{http://xxx.lanl.gov/abs/1602.02928}{{\normalfont
  [arXiv:astro-ph.GA/1602.02928]}}.
\newblock {\url{https://doi.org/10.3847/0067-0049/224/2/23}}.

\bibitem[{Ma{\'\i}z Apell{\'a}niz} \em{et~al.}(2020){Ma{\'\i}z Apell{\'a}niz},
  {Pantaleoni Gonz{\'a}lez}, {Barb{\'a}}, {Garc{\'\i}a-Lario}, and
  {Nogueras-Lara}]{maiz2020}
{Ma{\'\i}z Apell{\'a}niz}, J.; {Pantaleoni Gonz{\'a}lez}, M.; {Barb{\'a}},
  R.H.; {Garc{\'\i}a-Lario}, P.; {Nogueras-Lara}, F.
\newblock {Galactic extinction laws - I. A global NIR analysis with 2MASS
  photometry}.
\newblock {\em \mnras} {\bf 2020}, {\em 496},~4951--4963,
  \href{http://xxx.lanl.gov/abs/2006.09206}{{\normalfont
  [arXiv:astro-ph.GA/2006.09206]}}.
\newblock {\url{https://doi.org/10.1093/mnras/staa1790}}.

\bibitem[{Nogueras-Lara} \em{et~al.}(2019){Nogueras-Lara}, {Sch{\"o}del},
  {Najarro}, {Gallego-Calvente}, {Gallego-Cano}, {Shahzamanian}, and
  {Neumayer}]{Nogueras-Lara2019}
{Nogueras-Lara}, F.; {Sch{\"o}del}, R.; {Najarro}, F.; {Gallego-Calvente},
  A.T.; {Gallego-Cano}, E.; {Shahzamanian}, B.; {Neumayer}, N.
\newblock {Variability of the near-infrared extinction curve towards the
  Galactic centre}.
\newblock {\em \aap} {\bf 2019}, {\em 630},~L3,
  \href{http://xxx.lanl.gov/abs/1909.02494}{{\normalfont
  [arXiv:astro-ph.SR/1909.02494]}}.
\newblock {\url{https://doi.org/10.1051/0004-6361/201936322}}.

\end{thebibliography}

%=====================================
% References, variant B: internal bibliography
%=====================================

% If authors have biography, please use the format below
%\section*{Short Biography of Authors}
%\bio
%{\raisebox{-0.35cm}{\includegraphics[width=3.5cm,height=5.3cm,clip,keepaspectratio]{Definitions/author1.pdf}}}
%{\textbf{Firstname Lastname} Biography of first author}
%
%\bio
%{\raisebox{-0.35cm}{\includegraphics[width=3.5cm,height=5.3cm,clip,keepaspectratio]{Definitions/author2.jpg}}}
%{\textbf{Firstname Lastname} Biography of second author}

% For the MDPI journals use author-date citation, please follow the formatting guidelines on http://www.mdpi.com/authors/references
% To cite two works by the same author: \citeauthor{ref-journal-1a} (\citeyear{ref-journal-1a}, \citeyear{ref-journal-1b}). This produces: Whittaker (1967, 1975)
% To cite two works by the same author with specific pages: \citeauthor{ref-journal-3a} (\citeyear{ref-journal-3a}, p. 328; \citeyear{ref-journal-3b}, p.475). This produces: Wong (1999, p. 328; 2000, p. 475)

%%%%%%%%%%%%%%%%%%%%%%%%%%%%%%%%%%%%%%%%%%
%% for journal Sci
%\reviewreports{\\
%Reviewer 1 comments and authors’ response\\
%Reviewer 2 comments and authors’ response\\
%Reviewer 3 comments and authors’ response
%}
%%%%%%%%%%%%%%%%%%%%%%%%%%%%%%%%%%%%%%%%%%
\PublishersNote{}
\end{adjustwidth}
\end{document}